\documentclass[journal]{IEEEtran}

\ifCLASSINFOpdf
\else
\fi
\usepackage{amsmath,amssymb,amsfonts}
\usepackage[font=small]{caption}
\usepackage{algorithmic}
\usepackage{graphicx}
\usepackage{subcaption}
\usepackage{textcomp}
\usepackage{enumitem}
\usepackage{booktabs}
\usepackage{lipsum}
\usepackage{tabularx}
\usepackage{float}
\usepackage{placeins}
\usepackage{varioref}
\usepackage[utf8]{inputenc}
\usepackage{tikz}
\usepackage{mathtools,mathdots}
\usepackage[outline]{contour}
\usepackage{multirow}
\usepackage[style=ieee]{biblatex}
\usepackage[english]{babel}
\def\sina#1{\textcolor{black}{#1}}
\def\nrc#1{\textcolor{black}{#1}}

\def\nrcR#1{\textcolor{black}{#1}}

\begin{document}
\title{\nrc{An Approach for Fast Cascading Failure Simulation in Dynamic Models of Power Systems}}

\author{Sina Gharebaghi,~\IEEEmembership{Student Member,~IEEE,}
        Nilanjan Ray Chaudhuri,~\IEEEmembership{Senior Member,~IEEE,}
        Ting He,~\IEEEmembership{Senior Member,~IEEE,}
        and Thomas La Porta,~\IEEEmembership{Fellow,~IEEE}
         
\thanks{Authors are with The School of Electrical Engineering and Computer Science, The Pennsylvania State University, University Park,
PA 16802, USA. e-mail: svg5765@psu.edu, nuc88@engr.psu.edu, tzh58@psu.edu, and tfl12@psu.edu.}
\thanks{Financial support from NSF Grant Award ECCS 1836827 is gratefully acknowledged.}}
\markboth{}%
{Shell \MakeLowercase{\textit{et al.}}: Bare Demo of IEEEtran.cls for IEEE Journals}
\maketitle

\begin{abstract}
The ground truth for cascading failure in power system can only be obtained through a detailed dynamic model involving nonlinear differential and algebraic equations whose solution process is computationally expensive. This has prohibited adoption of such models for cascading failure simulation. 
To solve this, we propose a fast cascading failure simulation approach based on  implicit Backward Euler method (BEM) with stiff decay property. Unfortunately, BEM suffers from  hyperstability issue in case of oscillatory instability and converges to the unstable equilibrium. We propose a predictor-corrector approach to fully address the hyperstability issue in BEM.  \nrc{The predictor identifies oscillatory instability based on eigendecomposition of the system matrix at the  post-disturbance unstable equilibrium obtained as a byproduct of BEM. The corrector uses right eigenvectors to identify the group of machines participating in the unstable mode. This helps in applying appropriate protection schemes as in ground truth.} We use \nrc{Trapezoidal method (TM)-based simulation as the benchmark to validate the results of the proposed approach on the IEEE $118$-bus network, $2,383$-bus Polish grid,} \nrcR{and IEEE $68$-bus system.} The proposed approach is able to track the cascade path and replicate the end results of TM-based simulation \nrc{with very high accuracy} while reducing the average simulation time by $\approx10-35$ fold. \nrcR{The proposed approach was also compared with the partitioned method, which led to similar conclusions.}
\end{abstract}


\IEEEpeerreviewmaketitle

\section{Introduction}

\IEEEPARstart{C}{ascading} \nrc{failure study in highly complex dynamical systems like electric power grids is   challenging as it demands long-term simulations of models involving solutions of many nonlinear differential and algebraic equations.   
As a result, it is very difficult to perform statistical analysis of cascading failure using such models. This has led to application of less accurate but computationally manageable quasi-steady-state (QSS) models; see for example \cite{Henneaux_Benchmarking} for a comprehensive source of references. The objective of this paper is to propose an approach for fast cascading failure simulation that accurately traces the cascade path and lends itself to statistical analyses.}

\nrc{At the outset, we clarify that our goal is to perform deterministic cascading failure analysis, which implicitly assumes
that all systems act as expected during the cascade, i.e., potential mistripping of protective relaying and other malfunctions are not considered during the cascade \cite{Pierre_Dyn_Investment,dyn2_song}. This is in contrast to probabilistic approaches that consider
that the evolution of the power system after an initial set of
contingencies can follow multiple trajectories, see for example \cite{Henneaux_Dyn_Probab,Vallem_Dyn_Hybrid}.}
\vspace{-2pt}
\nrc{\subsection{Literature on Dynamic Simulation of Cascading Failure}\label{sec:CascadeLit}
Unlike their QSS counterpart, the literature on dynamic models of cascading failure is relatively limited -- see for example \cite{Flueck_Dyn_Protection,Parmer_Dyn_HPC,dyn3_Khaitan,Vallem_Dyn_Hybrid,Henneaux_Dyn_Probab,dyn2_song,Pierre_Dyn_Investment,Schafer-18-Dynamically_InducedCascade} and references therein. The papers can be broadly divided into three categories.}

\nrc{(1) \textit{Review- \& proposition-type papers:} For example, authors in \cite{Flueck_Dyn_Protection} present a brief review of existing modeling techniques and  simulation frameworks for cascading failure analysis, and discuss open questions related to interaction between protection systems and cascading failure. Authors in \cite{Parmer_Dyn_HPC} propose the development of a dynamic power system simulator that has the ability to tune the present direct linear solver, nonlinear solver, and the DAE integrator. In the same line, reference \cite{dyn3_Khaitan} suggests a parallelized algorithm for cascade simulations. The focus is to increase the simulation speed through parallelization strategy intended for deployment on a super computer.}

\nrc{(2) \textit{Papers proposing hybrid cascading failure models:}
Authors in \cite{Vallem_Dyn_Hybrid} developed a cascading failure simulation tool called dynamic contingency analysis tool (DCAT), which employs a hybrid approach of simulation that judges the stress of the system and switches between QSS and dynamic simulations. In addition to standard relay modeling, misoperations like stuck breakers are considered and corrective actions in post-transient steady-state conditions are included in the proposed model. 
}

\nrc{(3) \textit{Papers proposing dynamic cascading failure models:} Paper \cite{Henneaux_Dyn_Probab} proposed a two-Level probabilistic risk assessment
of cascading outages. Dynamic cascade events are separated into two categories, slow and fast cascade. 
The paper combines probabilistic simulations for the slow and the fast cascading events using different degree of details in the dynamic models.}

\nrcR{Schafer et-al~\cite{Schafer-18-Dynamically_InducedCascade} proposed to include network dynamics in the model to study dynamically-induced cascading failure. They represented the synchronous generator dynamics through swing equations. However, swing dynamics might not constitute an adequate representation of the synchronous machines as exciters play an important role in electromechanical oscillations \cite{kundur_book}.}

\nrc{Reference \cite{dyn2_song} proposed a detailed dynamic model for deterministic cascade propagation analysis. The method is tested with randomly selected $N-2$ contingencies. The authors conclude that the load model is very critical in evaluating the risk of cascading failures. It was also shown that the DC QSS model can reasonably approximate the cascade path in the early stages and deviates from the ground truth in later stages.}

\nrc{Paper \cite{Pierre_Dyn_Investment} proposed a multi-time period two-stage stochastic mixed-integer linear optimization model to specify the optimal investment on the network to enhance system's resilience against natural disasters. The model uses dynamic
simulations for cascading failure simulation, and the multi-time period restoration, modeled through a DC optimal power flow initialized by the solution of dynamic simulation.}
\vspace{-2pt}
\nrc{\subsection{Gaps in Literature}\label{sec:LitGaps}
The first category of papers \cite{Flueck_Dyn_Protection,Parmer_Dyn_HPC,dyn3_Khaitan} either reviewed the state-of-art or made propositions, but no cascading failure simulations were performed in these works.}

\nrc{Although references in the second \cite{Vallem_Dyn_Hybrid} and the third category \cite{Henneaux_Dyn_Probab,dyn2_song,Pierre_Dyn_Investment} have made valuable contributions, they still suffer from the computational burden faced by the simulation of dynamic models. For example, \cite{dyn2_song} effectively simulated $88$ cases in Polish system out of $1,200$ that can be called cascades because most ($1,081$) did not have any dependent outages (i.e., any further outages following the initial outages) leading to short simulations, while $31$ diverged. Hybrid simulation \cite{Vallem_Dyn_Hybrid} strategy can reduce simulation time, but may face accuracy issues as it is complicated to switch between dynamic and QSS simulations. At any rate, analysis in  \cite{Vallem_Dyn_Hybrid} starts with dynamic simulations -- hence the bottleneck remains.}

\nrc{The reason behind this is the fact that dynamic simulations in these works use a similar structure and the same integration methods as in the conventional planning models. The objective of traditional planning studies is to perform $N-1$ and $N-2$ contingency simulations that normally last up to $30$ s. They are computationally very expensive and not suitable for running cascading failure simulations.}
\vspace{-2pt}
\nrc{
\subsection{Contribution of Our Work}\label{sec:Contribution}
We propose a fast time-domain cascading failure simulation approach based on implicit Backward Euler method (BEM) with stiff decay property, in which large time-step can be used to speed up simulations. However, one disadvantage of BEM is the hyperstability issue in case of oscillatory instability that leads to convergence to the unstable equilibrium. We propose a predictor-corrector approach (PC-approach) to fully address the hyperstability issue in BEM. We also propose an adaptive center of inertia (COI) reference frame-based approach to ensure faster convergence of Newton iterations. Unlike the traditional COI frame-based method \cite{kundur_book,sauer_pai_book}, our approach seamlessly works for cascading failure simulation leading to island formation. The objective of our proposed method is to trace the exact cascade path during simulation and reproduce the exact end result of cascade with respect to the ground truth. We use a dynamic model which applies Trapezoidal method (TM) for numerical integration as a benchmark to test our proposed model for cascade simulations. Results on the IEEE $118$-bus system, \nrcR{IEEE $68$-bus system,} and the $2,383$-bus Polish network show high accuracy and significant speedup in simulation with multi-tier cascading failures.} \nrcR{We also show that the proposed approach maintains a significant speedup gain compared to the partitioned approach with an explicit numerical integration method.}

\section{Dynamic Simulation: Preliminaries \& State-of-art }\label{sec_II}
We first look into the structure of traditional dynamic simulation methods used for power system planning studies. Next, we elaborate on the challenges in using them for power system cascading failure simulation.

\subsection{Dynamic Simulation Preliminaries}\label{sec:DynPrelim} 
Power system's dynamic model is typically represented by a set of nonlinear differential algebraic equations (DAEs) \cite{kundur_book}. \nrcR{These equations can be represented in the following compact form after augmenting them with an \textit{implicit} discrete variable $z$ to model relay actions when the related constraint involving function $h\left(  \cdot  \right)$ is violated 
\begin{equation}
\begin{array}{l}\label{eq_x_dot}
\Dot{x} =  f(x,V,z)
\end{array}
\end{equation}
\begin{equation}
\begin{array}{l}\label{eq_Injec}
0 = I(x,V,z) - Y_N(z)V
\end{array}
\end{equation}
\begin{equation}
\begin{array}{l}\label{eq_Constraint}
0 \succ h(x,V,z).
\end{array}
\end{equation}}

\vspace{-6pt}
Here, $x \in \mathbb{R}^n$ is the state vector consisting of individual device states, $V \in \mathbb{R}^m$ denotes the vector of real and imaginary components of bus voltages, \nrcR{$z \in \mathbb{Z}^p$ is a discrete variable whose elements can assume values $0$ or $1$ indicating status of circuit breakers operated by relays}, $I \in \mathbb{R}^m$ constitutes of real and imaginary components of current injection phasors in buses, $ Y_N  \in \mathbb{R}^{m \times m} $ is the admittance matrix of network in its real form (i.e., separating the real and imaginary parts of the equations), \nrcR{and $h: \mathbb{R}^n\times\mathbb{R}^m\times \mathbb{Z}^p \to \mathbb{R}^q $ indicates line currents should be below their ratings and bus voltages below corresponding thresholds, among others. If the inequality constraint \eqref{eq_Constraint} is violated, the relevant relay will determine the trip time $T_{trip}$ and start a countdown process. When $T_{trip}$ becomes zero, the corresponding element of $z$, whose nominal value is $1$, also becomes $0$. This changes the $Y_{bus}$ and/or the injected current $I$. If the inequality constraint violation no longer holds before $T_{trip}$ goes to zero, the countdown stops. The details of different types of relay actions have been included in Section IV.A.} 

Dynamic simulation in power system solves an initial value problem (IVP) on the DAEs \eqref{eq_x_dot}, \eqref{eq_Injec}  with a set of known initial conditions \nrcR{$(x_0,V_0, z_0) \in \mathbb{R}^n\times \mathbb{R}^m\times \mathbb{R}^q$}. \textit{For a cascading failure simulation, such IVPs are solved repeatedly following each event, where an event refers to a discontinuity introduced by fault, line tripping, load shedding, and so on.}

\nrc{There are two philosophies for solving the IVPs in power systems literature -- partitioned and simultaneous \cite{kundur_book}.  In the partitioned approach, the algebraic and the differential equations are solved sequentially, whereas the simultaneous approach uses implicit integration methods, which combines them to pose them as a set of nonlinear algebraic equations. Commercial simulation softwares use second-order Adams-Bashforth (AB2) method~\cite{Juan-16-PV} - an explicit method - in the partitioned approach. \nrcR{ It is well-known, see for example page 861 of \cite{kundur_book}, that production-grade stability programs use partitioned approach because of programming flexibility, simplicity, reliability, and robustness. However, it has also been mentioned that its main drawback is numerical instability of explicit methods. \textit{Such methods are typically run at a fixed time step to avoid numerical instability issues. On the other hand, simultaneous approach with implicit integration methods can run with a variable time-step.} They are more widely explored in academic research~\cite{dyn2_song}, and is followed in our work.}  
}
\vspace{-2pt}
\subsection{Simultaneous Solution: State-of-art} \label{sec:SimultImplicit}	

\nrc{Here we briefly describe state-of-art on the simultaneous approach where perhaps the most popular implicit integration method is TM~\cite{dyn2_song}. \nrcR{In the context of solving DAEs described in the previous section, note that $z$ is an \textit{implicit variable} and does not appear explicitly in the numerical integration process, except that it brings in discontinuities. To avoid clutter, going forward we will drop $z$ from equations and will describe how discontinuities are handled later.} Discretization of \eqref{eq_x_dot} using TM results in the following expression}\vspace{2pt}
\begin{equation}
\begin{array}{l}\label{eq_TM}
F(x_{n+1},V_{n+1}) = x_{n+1} - x_n - \frac{\Delta t}{2} (f(x_{n+1},V_{n+1}) \\[4pt] ~~~~~~~~~~~~~~~~~~~~~~~~+ f(x_{n},V_{n})) 
\end{array}
\end{equation}
\nrc{where, $\Delta t$ is the step-size of integration, subscript $n$ corresponds to time instant $t_n$, and $F$ is the mismatch function for differential equations. The mismatch function for algebraic equations is defined as follows}
\begin{equation}
\begin{array}{l}\label{eq_mismatch_alg}
G(x_{n+1},V_{n+1}) = Y_N V_{n+1} - I(x_{n+1},V_{n+1})
\end{array}
\end{equation}
\nrc{where, $x_{n+1}$  and $V_{n+1}$ are found by simultaneously solving the following nonlinear algebraic equations}
\begin{equation}
\begin{array}{l}\label{eq_Feq0}
F(x_{n+1},V_{n+1}) = 0, ~~
G(x_{n+1},V_{n+1}) = 0.
\end{array}
\end{equation}

\nrc{Typically, Newton's method \cite{Petzold_book} is used for solving these equations. For the $(k+1)^{th}$ iteration of Newton's method, we have}\vspace{2pt}
\begin{equation}
\begin{array}{l}\label{eq_var_upd}
\begin{bmatrix} 
x_{n+1}^{k+1}\\[6pt]
V_{n+1}^{k+1}\\
\end{bmatrix}
=
\begin{bmatrix}
x_{n+1}^{k}\\[6pt]
V_{n+1}^{k}\\
\end{bmatrix}
 + 
\begin{bmatrix}
\Delta x_{n+1}^{k}\\[6pt]
\Delta V_{n+1}^{k}\\
\end{bmatrix}
\end{array}
\end{equation}
\vspace{3pt}
\begin{equation}
\begin{array}{l}\label{eq_inverse_jac}
\begin{bmatrix} 
\scalebox{0.90}{$
-F(x_{n+1}^k,V_{n+1}^k)$}\\[6pt]
\scalebox{0.90}{$-G(x_{n+1}^k,V_{n+1}^k)$}\\
\end{bmatrix}
=
\begin{bmatrix}
\frac{\partial F}{ \partial x_{n+1}}   &  \frac{\partial F}{ \partial V_{n+1}} \\[6pt]
\frac{\partial G}{ \partial x_{n+1}}   &  \frac{\partial G}{ \partial V_{n+1}}\\
\end{bmatrix}_{n+1}^k
\begin{bmatrix}
\Delta x_{n+1}^{k}\\[6pt]
\Delta V_{n+1}^{k}\\
\end{bmatrix}
\end{array}
\end{equation}
\vspace{4pt}
\begin{equation}\label{eq_jacob}
\begin{bmatrix}
J \\
\end{bmatrix}
=
\begin{bmatrix}
\frac{\partial F}{ \partial x_{n+1}}   &  \frac{\partial F}{ \partial V_{n+1}} \\[6pt]
\frac{\partial G}{ \partial x_{n+1}}   &  \frac{\partial G}{ \partial V_{n+1}}\\
\end{bmatrix}_{n+1}^k
=
\begin{bmatrix}
J_{11}   &  J_{12} \\[6pt]
J_{21}   &  J_{22}\\
\end{bmatrix}
\end{equation}
\nrc{where, $J$ is the Jacobian matrix. First,  $\Delta x$ and $\Delta V$ are calculated using (\ref{eq_inverse_jac}), which in turn are used to update $x$ and $V$  through (\ref{eq_var_upd}). Newton iterations are stopped when $||[F^T ~G^T]^T||_{\infty} \leq \epsilon$, where $\epsilon \in \mathbb{R_+}$ is the tolerance for convergence. }

\nrc{\textit{\underline{Remarks on state-of-art:}}\\
1. \textit{Variants of Newton iterations:} 
Three popular variants are full Newton's method, dishonest/very dishonest Newton's method (VDHN) and quasi-Newton method \cite{EPRI_report_power_sys_dyn_ana}.\\
2. \textit{Jacobian calculation:} 
\sina{Both direct analytical method and difference approximation method \cite{Petzold_book} have been used. 
}\\
3. \textit{Solution of linear equation $Ax = b$:} Since the Jacobian is very sparse, solving the set of linear equations \eqref{eq_inverse_jac} in the general form $Ax = b$ takes significant advantage of this aspect during storage and computations. Both direct solution methods like sparsity-oriented triangular factorization~\cite{Tinney-73-Triangular} and KLU~\cite{Davis-04-KLU}, and iterative solutions like Preconditioned Conjugate Gradient (PCG)~\cite{ortega1988introduction,Vorst-90-Siam}, and General Minimal Residual (GMRES) method~\cite{Vorst-90-Siam} have been proposed. \\
4. \textit{Variable time-step:} \sina{To speed up the simulation, adaptive time step size control is used based on local truncation error (LTE) \cite{Griffiths_IVP_book} that leads to large time steps when solution is not varying rapidly.}
\\
5. \textit{Suitability for cascading failure simulation:} Even applying sparse computations for solving \eqref{eq_inverse_jac} and using a variable time-step TM-based solver, the state-of-art suffers from significant computational burden during cascading failure simulation, since they take much longer than typical $N-1$ or $N-2$ contingency simulations that last $30$ s or less.}
\\
\nrcR{6. \textit{Handling discrete events:} If the integration time step $\Delta t$ calculated by the variable-step algorithm is more than the time remaining before $T_{trip}$ becomes zero, then $\Delta t$ is truncated to match the tripping instant. When $T_{trip}$ becomes zero, at $t=t_n$, a discrete event occurs due to relay action, i.e., certain elements of $z$ becomes $0$. In this case, the following steps are performed -- (a) network configuration is updated and the corresponding $Y_{bus}$ is calculated; (b)  $V_n$ is updated by iteratively solving \eqref{eq_mismatch_alg} for $t = t_n$. To that end, $Y_N$ is updated if needed and the $J_{22}$ sub-matrix of the Jacobian in \eqref{eq_jacob} is used. Next, updated $x_n$ and $V_n$ are used as the initial guess for $t = t_{n+1}$, i.e., $\left[ {\begin{array}{*{20}c}
   {\left( {x_{n + 1}^0 } \right)^T } & {\left( {V_{n + 1}^0 } \right)^T }  \\
\end{array}} \right]^T  = \left[ {\begin{array}{*{20}c}
   {\left( {x_n } \right)^T } & {\left( {V_n } \right)^T }  \\
\end{array}} \right]^T $ is used. Equation \eqref{eq_inverse_jac} is then iteratively solved to find $\left[ {\begin{array}{*{20}c}
   {x_{n + 1}^T } & {V_{n + 1}^T }  \\
\end{array}} \right]^T $; (c) the time-step is reduced to the minimum step-size of $\Delta t_{min}$ for a pre-defined period of simulation; (d) the IVPs based on the predicted initial conditions are solved using the reduced time-step. }

\nrc{Going forward, we define `ground truth' as the cascading failure simulation results produced by a benchmark model that uses (a) variable-step TM~\cite{dyn2_song} with $\Delta t \in [0.002,1]$ s, $\epsilon = 10^{-4}$ that leverages an adaptive COI reference frame-based approach described in Section~\ref{sec:COI_frame_models}, (b) formulates the Jacobian analytically, (c) applies full Newton iterations, (d) uses sparse objects for storage and calculations, and (e) applies Matlab's most comprehensive inversion routine for solving \eqref{eq_inverse_jac}, see flowchart in \cite{matlab_mldivide}. The model \nrcR{consists of $4^{th}$-order synchronous generator model} equipped with the same \nrcR{governors, static exciters, and} relays as our proposed model described in the next section, except that the special protection scheme (SPS) is not functional, but measurement-based. The benchmark and the proposed models are built from the first principles in Matlab \cite{Matlab_citation} and MATPOWER \cite{MATPOWER} is used for power flow solution used during initialization. } \sina{Simulation is stopped if i) speed variation of machines in a predetermined window length is below a certain threshold, and no future relay actions are anticipated, or ii) a complete collapse is observed.} 

\section{Proposed Methodology for Dynamic Simulation of Cascading Failure} \label{sec_III}	

At the outset, we define what is expected out of a dynamic cascading failure simulation model --
\begin{enumerate}
    \item The model should be able to capture the exact cascade propagation path as in ground truth.
    \item The model should give exact end-result of cascade as the ground truth in terms of topology, voltage profile, frequency, and demand served.
    \item The model should be computationally efficient, so that statistical analyses can be performed, which is critical for cascading failure studies.
\end{enumerate}
Even though it would be ideal if the dynamic model is able to simulate the exact trajectories of state and algebraic variables of the system as the ground truth and also lends itself to statistical analyses -- unfortunately, that has proven to be elusive thus far \cite{Pierre_Dyn_Investment,dyn2_song,Henneaux_Dyn_Probab,Vallem_Dyn_Hybrid,Flueck_Dyn_Protection,Parmer_Dyn_HPC,dyn3_Khaitan,Schafer-18-Dynamically_InducedCascade}. \textit{We argue that if the above objectives are met at the expense of accurate tracking of trajectories of system variables, it  should be sufficient for dynamic cascading simulations without compromising accuracy of statistical analyses. } 

To that end, we propose the following (see, Fig.~\ref{fig:fig_flow_chart_predictive_corrective}) --\\
\textbf{(a)} A time-domain simulation approach based on a stiff decay integration method. More specific, we apply implicit backward Euler method (BEM) \cite{Petzold_book} for the simultaneous solution process.\\
\textbf{(b)} We solve the hyperstability issue of BEM using an eigen analysis-based \textit{predictor-corrector} method, which leverages its stiff-decay and hyperstability properties.\\
\textbf{(c)} We propose \textit{functional} implementation of SPS against unstable interarea oscillations \nrc{and generator non-first swing out-of-step protection (e.g., due to an unstable local mode).}\\
\textbf{(d)} We model time-delayed overcurrent (OC), local undervoltage load shedding (UVLS), and generator first swing out-of-step relays. Note that other types of relays can also be modeled in the proposed framework.\\
\nrc{\textbf{(e)} We propose an adaptive COI frame-based approach that can seamlessly work during island formation.}

\subsection{BEM: Absolute Stability and Stiff Decay Properties \cite{Petzold_book}} \label{sec:BEM}	
BEM is derived using a Taylor expansion centered at $t_{n+1}$, which is a first-order method 
\cite{Petzold_book}. Discretizing (\ref{eq_x_dot}) using BEM results in the following expression
\begin{equation}
\begin{array}{l}\label{eq_BEM}
F(x_{n+1},V_{n+1}) = x_{n+1} - x_n - {\Delta t} f(x_{n+1},V_{n+1})
\end{array}
\end{equation}

\subsubsection{Absolute stability property}\label{sec:AbsStab}
First, we analyze the absolute stability property of TM and compare it with that of BEM. The standard approach for this is to consider the so-called \textit{test equation} $\Dot{x} =  \lambda x$, where $\lambda$ is a complex number denoting the eigenvalue of a system matrix. The \textit{region of absolute stability} is defined as the region in the complex  $\lambda\Delta t$-plane such that applying the numerical integration method for the test equation from within this region yields an approximate solution satisfying the absolute stability requirement $|x_{n+1}| \leq |x_{n}|$. By discretizing the test equation using TM, we have
\begin{equation}
\begin{array}{l}\label{eq_test_sys_TM}
x_{n+1} = \frac{2+\lambda\Delta t}{2-\lambda\Delta t} x_n;  ~~~ AF = |\frac{2+\lambda\Delta t}{2-\lambda\Delta t}|
\end{array}
\end{equation}
\noindent where, $AF$ is called amplification factor. Therefore, the region of absolute stability of TM can be obtained by the region that is satisfying $AF \leq 1$, which is the left half of the $\lambda\Delta t$ plane. Similarly, applying BEM to the test equation results in:
\begin{equation}
\begin{array}{l}\label{eq_test_sys_BEM}
x_{n+1} = \frac{1}{1-\lambda \Delta t} x_n;  ~~~ AF = |\frac{1}{1-\lambda \Delta t}|
\end{array}
\end{equation}

\begin{figure}[!t]
\centerline{\includegraphics[scale = 0.5, trim= 1.2cm 0.7cm 0.7cm 0.8cm, clip=true]{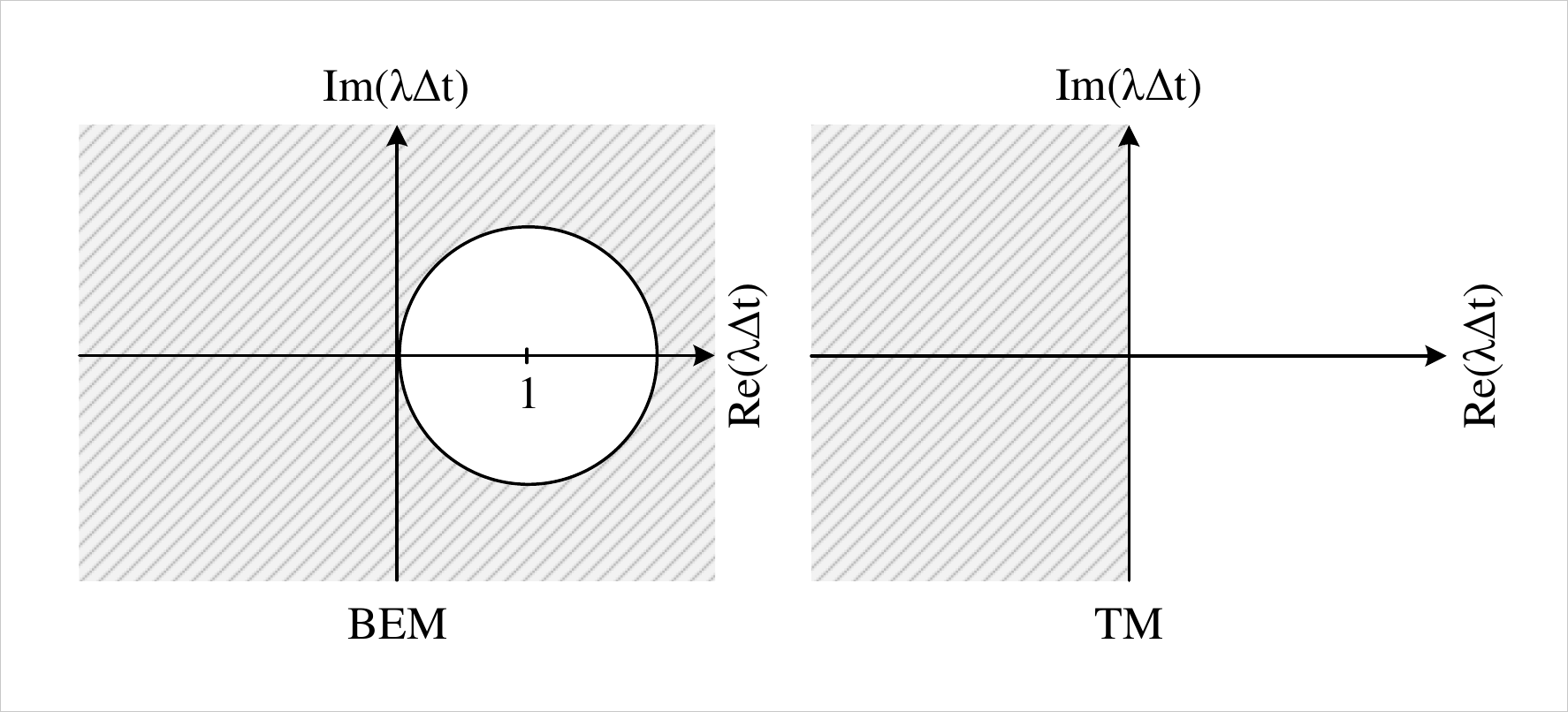}}
\caption{Absolute stability regions of BEM (left) and TM (right) shown in gray.}
\centering
\label{fig:absolute_stability_BEM_TM}
\end{figure}

Therefore, the region of absolute stability of BEM is the entire left half  of the $\lambda\Delta t$ plane in addition to the entire right half plane outside the unit circle centered at $(1,0)$. As shown in Fig. \ref{fig:absolute_stability_BEM_TM}, the regions of absolute stability in gray indicates that both TM and BEM are numerically \textit{A-stable}  \cite{Petzold_book}.

\subsubsection{Stiff decay property}\label{sec:StiffDecay}
In line with our argument presented earlier, in the dynamic simulation of cascading failure, one might not be interested in detailed transient oscillatory behavior of the system as long as the expectations are met. In this regard, using large time steps would be desired. However, the integration method should be robust enough to tolerate the large steps. According to \cite{Petzold_book}, when $\Re (\lambda\Delta t) \to -\infty$, for BEM we have $\frac{1}{1-\lambda \Delta t} \to 0$, however, for TM we have $\frac{2+\lambda\Delta t}{2-\lambda\Delta t} \to -1$. This property in BEM is called \textit{stiff decay}, representing ability of BEM in taking large steps to ignore fast oscillations in the dynamic model. On the other hand, one should not expect TM to act like integral methods with stiff decay property. 
This is due to the fact that the fast mode components of local errors for large time steps get propagated throughout the simulation interval \cite{Petzold_book}.

\subsubsection{Hyperstability issue of BEM}\label{sec:Hyperstab}
It is clear that the stiff decay property of BEM can be used to our advantage for dynamic simulations of cascading failure as it helps us take large time steps for ignoring fast oscillations and obtaining a coarse picture of the desired trajectories. However, BEM is never used in dynamic simulations of power system due to the \textit{hyperstability} problem \cite{Fabozzi_paper_BEM}. 

\nrcR{When a numerical integration method solves the differential equations of an unstable system and produces a stable response, then such a problem is called Hyperstability.}
This can be viewed from the absolute stability region of Fig. \ref{fig:absolute_stability_BEM_TM} for BEM satisfying $\Re \left( {\lambda \Delta t} \right) > 0$ and $\left( {\Re \left( {\lambda \Delta t} \right) - 1} \right)^2  + \left( {\Im \left( {\lambda \Delta t} \right)} \right)^2  > 1$. \nrcR{It  corresponds to the right half plane  outside the unit circle of the left subfigure. }  The practical implication of this is that BEM is not able to diagnose oscillatory instability if $\lambda \Delta t$ satisfies the above constraints. 

\begin{figure}[!t]
    \centering
    \begin{subfigure}{0.24\textwidth}
        \centering
        \includegraphics[scale = 0.095, trim= 0.85cm 0.5cm 4.8cm 0.5cm, clip=true]{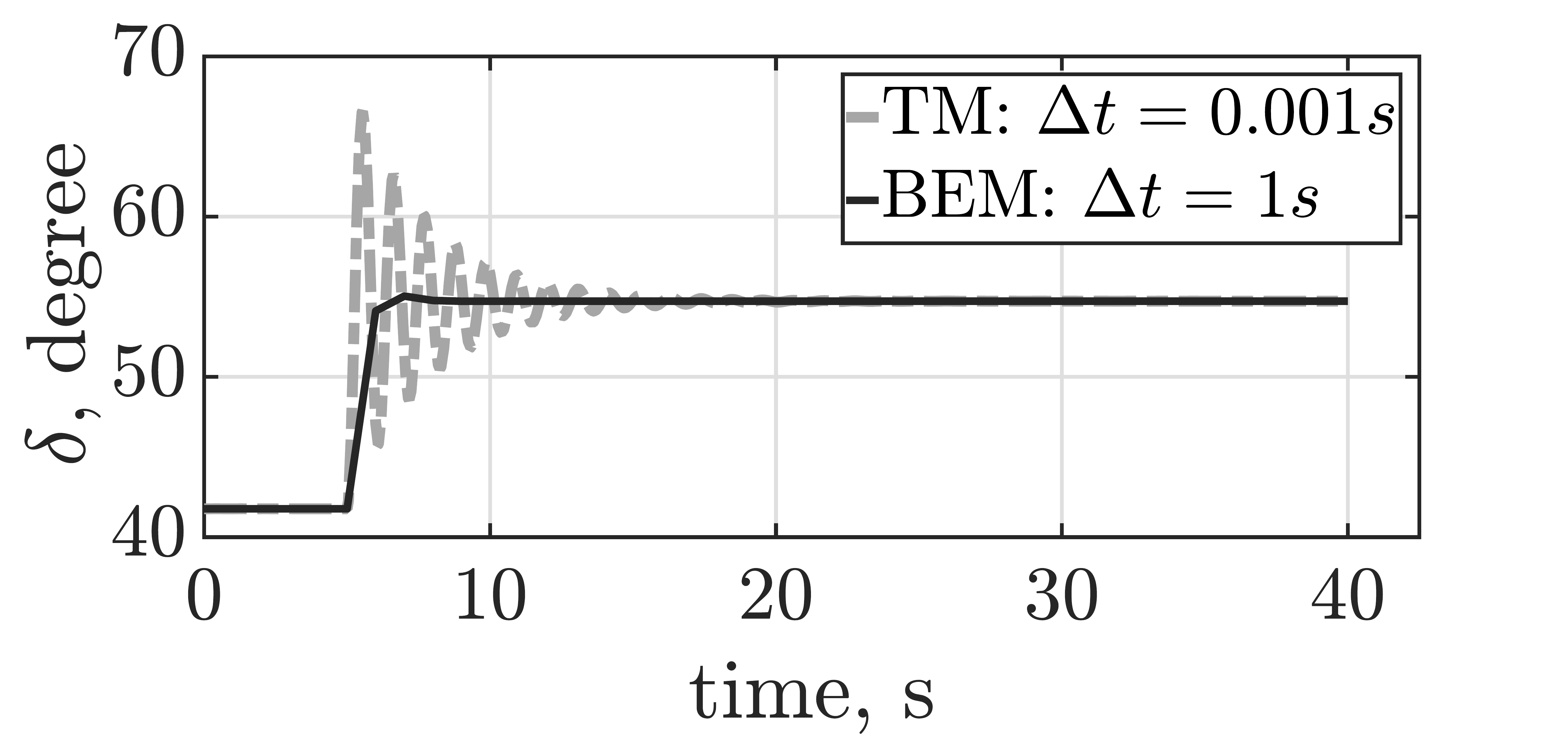}
    \end{subfigure}%
    \begin{subfigure}{0.24\textwidth}
        \centering
        \includegraphics[scale = 0.095, trim= .3cm 0.5cm 4.2cm 0.5cm, clip=true]{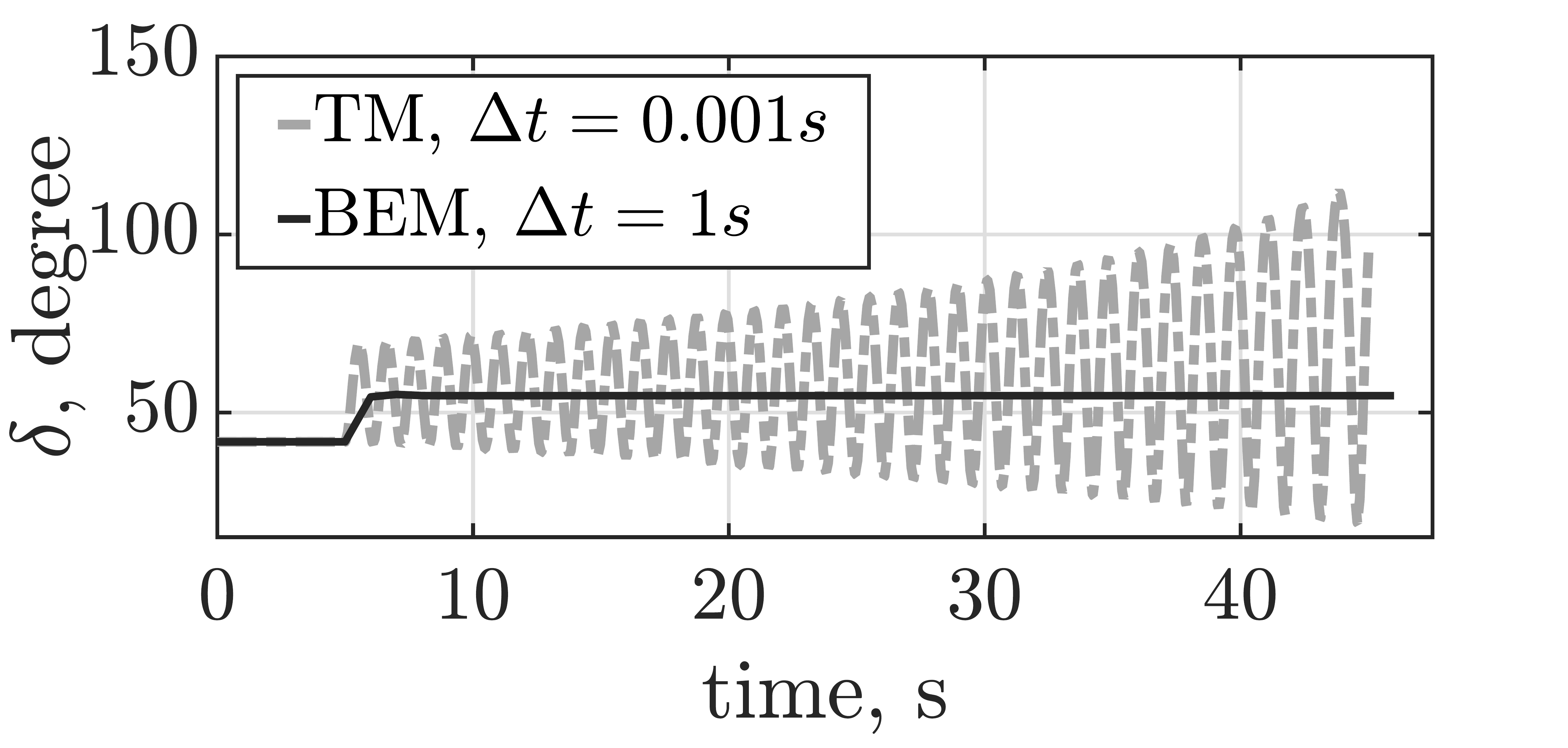}
    \end{subfigure}
    \caption{{\color{black} Rotor angle time-domain plot in SMIB system after one line outage at $t=5$ s; BEM vs TM in Left: Stable case, Right: \nrc{Oscillatory} instability case (hyperstability problem of BEM).}}
    \label{fig:SMIB_figures}
\end{figure}

Figure \ref{fig:SMIB_figures} compares the performance of BEM and TM in a single machine (represented by classical model) infinite bus (SMIB) system \cite{kundur_book} after tripping one of the double-circuit lines at $t=5$ s. The left and the right subplots represent a stable and an unstable case, respectively -- the latter is simulated by a negative damping factor. In both the scenarios, traditional model with TM is simulated with $\Delta t = 0.001$ s, and BEM uses much larger time step of $1$ s. It can be seen that for the stable case, the stiff decay property allows BEM to obtain the exact final result as TM while producing a coarse trajectory. For the unstable case however, the hyperstability problem of BEM is evident, where it converges to the unstable equilibrium point. Next, we will address the hyperstability problem of BEM in detail.

\begin{figure*}[h]
\centerline{\includegraphics[scale = 0.45, trim= 0.1cm 0.4cm 0.3cm 0.62cm, clip=true]{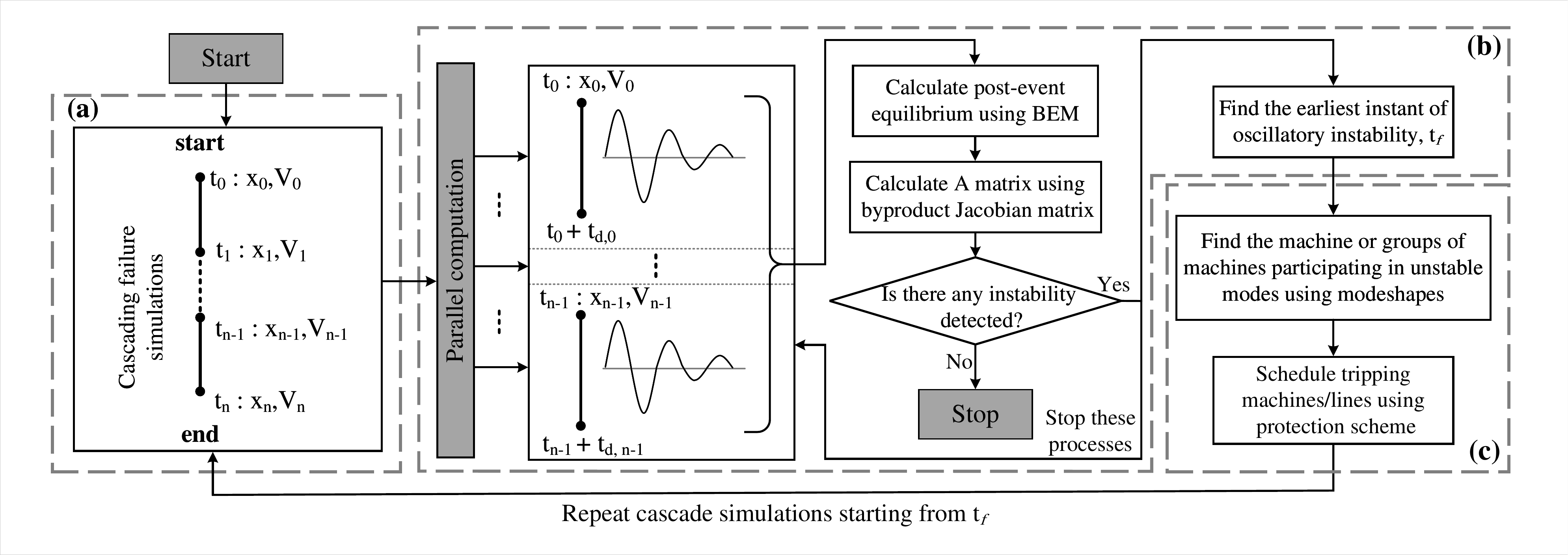}} 
\caption{Flowchart of proposed BEM with predictor-corrector (BEM-PC) approach. $t = t_i$: instants of tiers of cascade. Processes that can be run using parallel processors are indicated. (b): Predictor subprocess. (c): Corrector subprocess.}
\centering
\label{fig:fig_flow_chart_predictive_corrective}
\end{figure*}

\vspace{3pt}
\subsection{Addressing Hyperstabibility Problem of BEM using Predictor-Corrector Approach} \label{sec:PCapproach}

We propose a predictor-corrector (PC) approach to tackle the hyperstability problem in BEM, which is shown in a flowchart in Fig.~\ref{fig:fig_flow_chart_predictive_corrective}. In this flow chart, there are four key functions that are being performed in a \textit{serial-parallel} process.

\subsubsection{Cascading Failure Simulation Subprocesses \textbf{(a)}}\label{sec:(a_d)}
\nrc{In this subprocess, we run the cascading failure simulation using variable-step BEM, where OC, UVLS, and generator first swing out-of-step relays are modeled. The stopping criterion for BEM is similar to TM as described earlier, \nrcR{i.e., simulation is stopped if i) speed variation of machines in a predetermined window length is below a certain threshold and no future relay actions are anticipated, or ii) a complete collapse is observed.}} 
\sina {For step-size control in BEM, we use a different method than TM, }
\begin{equation}
\begin{array}{l}\label{eq_var_time_step_BEM}
\Delta t_{n+1} = \Delta t_{n} \frac{\tau}{\|F_x^0\|_\infty};~~\Delta t_k \in [\Delta t_{min},~ \Delta t_{max}]
\end{array}
\end{equation}
\noindent  \sina{where, ${\|F_x^0\|_\infty}$ shows the largest component of the first mismatch vector and $\tau$ is a hyperparameter to be tuned, see \cite{Fabozzi_paper_BEM} for explanation.} \nrc{Note that immediately following each event, we run simulation with $\Delta t = 0.002$ s for a pre-determined $k$ steps. This ensures that the non-oscillatory instability is captured. Moreover, if Newton iterations take more than $r$ iterations for convergence at a time instant, then we decrease time-step to $\Delta t = 0.002$ s.}


During the course of such a simulation, the instants of  tiers (events that alter the topology of the system) of cascade are marked by time variable $t = t_i$ in Fig.~\ref{fig:fig_flow_chart_predictive_corrective}. Corresponding to each such instant, we get the values of $x$ and $V$ as $[t_i: x_i,~V_i]$. Note that this subprocess runs in a \textit{serial} manner to solve a sequence of IVPs described in Section~\ref{sec_II}.


Following each such instant, there could be four broad types of unstable scenarios so far as voltage, angle, and frequency stability are concerned -- (1) local voltage instability, (2) frequency instability, (3) non-oscillatory angle instability, and (4) \nrc{local/interarea} oscillatory angle instability. Except the last phenomena, BEM does not face issues in capturing the others.

As discussed earlier, due to the hyperstability issue, BEM may converge to the unstable equilibrium following the fourth category of instability. This in turn can deviate the cascade propagation path from ground truth. \nrc{\textit{Our goals are to identify the earliest tier of onset of such instability, and execute appropriate protective actions that will be taken in the ground truth.}}

\subsubsection{Predictor Subprocess \textbf{(b)}}\label{sec:(b)}
\nrc{This subprocess shown in Fig.~\ref{fig:fig_flow_chart_predictive_corrective} runs after subprocess \textbf{(a)} ends. It constitutes running multiple simulations and calculations that are independent and thus \textit{parallelizable}.} The following steps are taken -- 
\begin{enumerate}[label=\roman*.]
    \item \textit{Solving independent IVPs for short duration:}  \sina {As subprocess \textbf{(a)} spits out $[t_i: x_i,~V_i]$ data, we solve IVPs with initial values $(x_i,~V_i)$ \nrc{that can be run} in the $i$th parallel processor using \nrc{variable-step} BEM. 
    \nrcR{ The simulation for the $i$th independent IVP is stopped if the speed variation of machines in a predetermined window length is below a certain threshold. As shown in Fig.~\ref{fig:fig_flow_chart_predictive_corrective}, $t_{d,i}$ is the time elapsed since the beginning of such a simulation when this stopping criterion is met.} In this period, we do not consider any event including relay actions.}
    
   \item \textit{Calculate system matrix for model linearized around post-event equilibrium:} \sina {Solving variable-step BEM within each parallel processor allows the trajectories to reach the post-event equilibrium --  \textit{\textbf{more critical, the post-event unstable equilibrium point due to BEM's stiff-decay and hyperstability properties}}. \nrc{We calculate the system matrix ($A$ matrix) of the model linearized around this equilibrium} as a \textit{byproduct} of BEM-based simulation using the elements of the Jacobian matrix as follows}\vspace{2pt}
    
    \begin{equation}\begin{array}{l}\label{eq_A_matrix}
    A = P_{11} + P_{12}P_{22}^{-1}P_{21}
     \end{array}
    \end{equation}
    
    
    
    \noindent where,
    
    \begin{equation}
    \begin{array}{l}\label{A_matrix_elements}
    P_{11} = \frac{1 }{\Delta t} (I-J_{11}); ~~
    P_{12} = -\frac{1 }{\Delta t} J_{12};\\[6pt]
    P_{21} = -J_{21}; ~~
    P_{22} = J_{22}
    \end{array}
    \end{equation}
\sina {
\noindent where, $I$ denotes the identity matrix and $\Delta t$ is the time step at the \nrc{end of duration $t_{di}$}. }

\item \textit{Eigendecomposition of $A$ matrix:}  \nrc{Eigendecomposition of  $A$ matrix is performed to detect oscillatory instability. For large systems, one can use selective unstable eigenvalue and corresponding right eigenvector calculations using the $S$-method \cite{Uchida-88-Smethod} or refer to relatively recent works on this topic \cite{BEZERRA2019113,Nelson-10-RightmostEigenvalue}.  The earliest event and corresponding time instant, say $t_f$ is identified at which the instability occurs. }

\end{enumerate}

\subsubsection{Corrector Subprocess \textbf{(c)}}\label{sec:(c)}
\nrc{If any oscillatory instablility is detected, its origin can be found from participation factors \cite{kundur_book}. Assuming the typical case of instability from electromechanical modes, we find the machine or groups of machines participating in these modes using their speed modeshapes calculated in subprocess \textbf{(b)}.  We schedule \textit{functional implementation} of the pre-determined protective action (e.g. out-of-step generator tripping or SPS action) that will take place following $t = t_f$ in the ground truth after a designed delay.}

\subsubsection{Restart subprocess \textbf{(a)}}\label{sec:Restart(a, d)}
As shown in Fig.~\ref{fig:fig_flow_chart_predictive_corrective}, with the knowledge of pre-determined protection action to be taken, we re-initiate the solution of IVPs at $t = t_f$ with initial states $x_f,~V_f$ and perform protection actions after a pre-defined delay and continue solving the subsequent IVPs.  

The above steps will be repeated until no instability is detected in the predictor subprocess. \vspace{-5pt}

\nrcR{
\subsection{Discussion on Computational Efficiency of BEM-PC}\label{sec:CompuEff}
In this Section, we present a brief \textit{qualitative} discussion on the computational efficiency of BEM-PC compared to TM using \textit{logical arguments}. Since it is difficult to analytically quantify this advantage, we have performed exhaustive comparison between the proposed method and TM through statistical analysis of CPU time needed for different subprocesses within BEM-PC in Section \ref{sec:CompuEffNumerical}.}

\nrcR{Our logical argument relies on the following points --
\begin{enumerate}
    \item Subprocess (a) in Fig.~\ref{fig:fig_flow_chart_predictive_corrective} runs with variable time algorithm \eqref{eq_var_time_step_BEM}, which allows significantly larger step size $\Delta t$ compared to that allowed in LTE-based variable step TM mentioned before. This is possible due to the \textit{stiff-decay} property of BEM as described in \ref{sec:BEM}. As a result, this subprocess can run much faster than TM. We have presented a statistical analysis of CPU time for running this subprocess compared to TM in Section \ref{sec:CompuEffNumerical}. In addition, we have also shown variation of time-step $\Delta t$ for TM and BEM in a few typical cases.
    \item Subprocess (b) is \textit{parallelizable}. It has three main tasks:\\
    \textbf{(b1)} Calculation of the post-event equilibrium,\\
    \textbf{(b2)} Calculation of $A$ matrix from \eqref{eq_A_matrix}, and\\ 
    \textbf{(b3)} Eigendecomposition (i.e., calciulation of eigenvalues and modeshapes) of $A$ matrix.\\
    The first part can be performed extremely fast with large $\Delta t$. The $A$ matrix calculation is a by-product and needs inversion of $J_{22}$, which is a \textit{highly sparse} matrix. We leverage sparse computation for this. Also, there are highly efficient routines for eigendecomposition. Further, note that once any unstable mode is found, $A$ matrix does not need to be calculated for the remaining equilibria.  We have demonstrated statistical analysis of CPU time needed for these individual steps in Section \ref{sec:CompuEffNumerical} \textit{without parallelization.}
    \item Subprocess (c) needs minimal computation as it is based on lookup table for enacting SPS action.
    \item Clearly, the computational efficiency of BEM relies on the fact that for a typical power system, a relatively small fraction of cascade simulations will lead to oscillatory instability, and therefore needs to re-run subprocess (a). 
\end{enumerate}
}

\nrcR{
The results in Section \ref{sec:CompuEffNumerical} support the above-mentioned arguments. In addition, a comparison with partitioned approach with fixed time-step-based explicit integration leads to similar conclusions.
}\vspace{4pt}

\nrcR{\textit{\underline{Remarks:}} \begin{enumerate}
    \item At a fundamental level, the proposed approach will be able to speed up any transient stability simulation of power systems with accurate end results. However, due to short-term nature of typical transient stability simulations the speed gain would be rather limited. 
    \item In contrast, cascading failure simulations may run for a much longer time, lead to formation of multiple islands, and show response across different time-scales throughout the process. BEM can run with a larger integration time-step than TM during most of the simulation period, which makes it ideally suited for longer term simulations. This argument becomes even more relevant since the proposed BEM-PC approach requires additional computations due to the prediction and the correction steps.
\end{enumerate}
}

\section{Modeling and Adaptive COI-frame-based Approach} \label{sec:COI_frame_models}
\subsection{Component and Relay Action Models}\label{sec:ComponentMdls}
We consider a $4^{th}$-order synchronous generator model (states $E_q^{'}$, $E_d^{'}$, $\delta$, $\Delta\omega$) with a first-order governor and static exciter models \cite{sauer_pai_book}. Both static constant power and dynamic loads in the form of synchronous condensers were considered. The synchronous condensers have similar models as generators, except that they do not have governors. 

\nrc{We have considered certain relay actions in our model. For example, undervoltage load shedding (UVLS) relays are associated with individual  buses connected to static loads, measuring an average voltage magnitude in a window of length $T_w^{UVLS}$ s. The relay trips $\lambda$ fraction of load if the avergae voltage magnitude of bus stays below threshold $V_{th}$ for $T_{tp}^{UVLS}$ s. The maximum number of times the UVLS relays are allowed to shed a specific load is $K_{max}^{shed}$.}

\nrc{The overcurrent (OC) relays measure an average magnitude of current flow in the lines in a $T_w^{OC}$ s window. The trip delay for an overloaded line is $T_{tp}^{line} = \frac{0.14}{(\frac{|I|}{I_c})^{0.02} - 1}$
where, $|I|$, and $I_c$ are average current flow in the present window and line heating limit, respectively. The window for OC relays is updated once in every second. Due to probable large amplitude oscillations in the line flows immediately following an event, OC relays use the latest pre-event trip delays till $1$ s following the event and then starts updating it. In addition, generator out-of-step relay action trips a machine with non-oscillatory instability. We have considered a specific type of pre-designed SPS action on oscillatory instability involving multiple machines as described in Section~\ref{sec:HyperstabStudy}. For both TM and BEM-PC, if the time remaining till the earliest scheduled trip instant by relays, $\Delta t_{trip}^{relay}$ is less than $\Delta t_{n+1}$ suggested by variable-step algorithms, we consider $\Delta t_{n+1} = \Delta t_{trip}^{relay}$.} 


\nrcR{\textit{\underline{Remarks on Modeling:}}\\ 1. Both BEM-PC and TM solve IVPs of the same DAEs, but the former can reach the post-disturbance equilibrium faster. This is possible because BEM-PC has stiff decay property that enables it to use a variable integration time-step with a larger step-size than TM during most of the simulation period.  \\
2. In certain cases the cascading process may include mid/long term stability issues involving aspects like boiler dynamics and long-term frequency instability. This however can easily be integrated in our proposed framework, and is not a limitation. Due to its stiff-decay property, BEM is ideally suited for longer term simulations and gives more benefit.
}

\subsection{Adaptive COI-frame-based Approach}\label{sec:AdaptiveCoi}
\begin{figure}[!b]
\centerline{\includegraphics[scale = 0.65, trim= 0.9cm 0.7cm 1.1cm 2.6cm, clip=true]{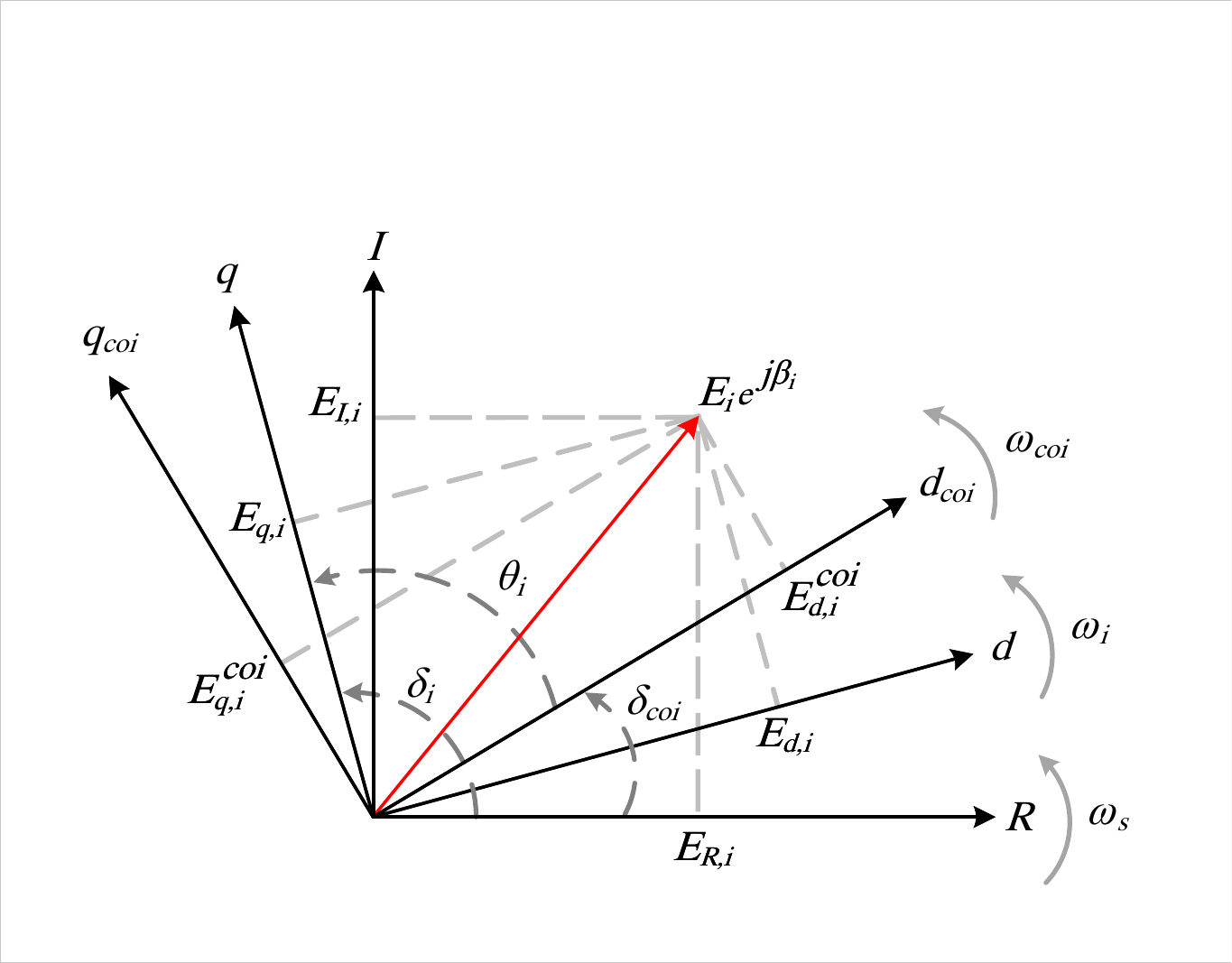}}
\caption{Three different reference frames.} 
\centering
\label{fig:RI_dq_coi_quad}
\end{figure}

\begin{figure}[!b]
\centerline{\includegraphics[scale = 0.95, trim= 0.45cm 0.6cm 0.3cm 0.5cm, clip=true]{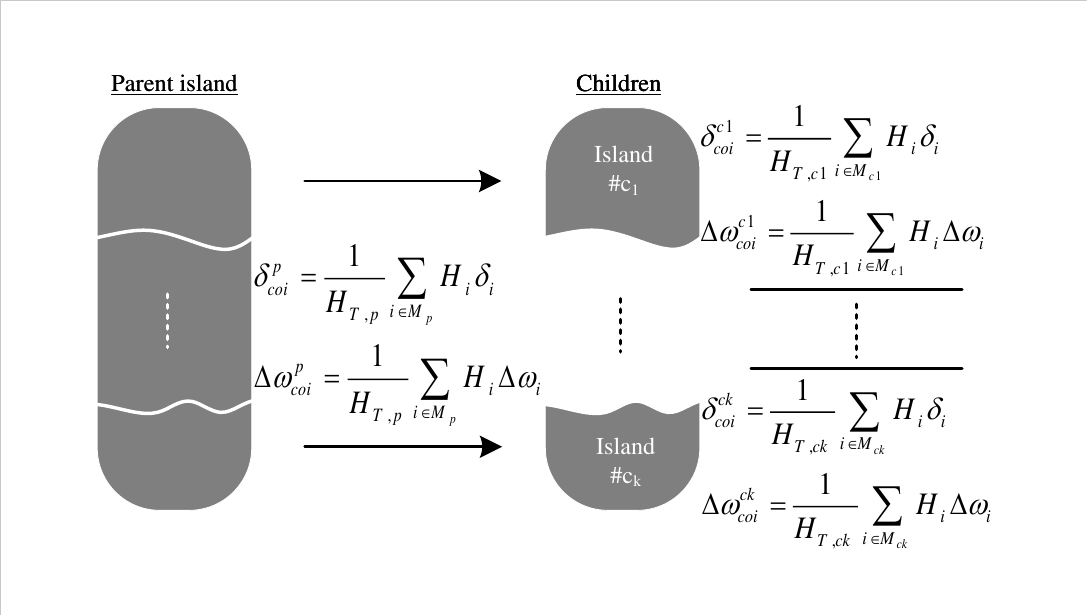}}
\caption{Island formation during cascading failure: superscripts $p$ for parent, $ci$ for $i$th child. } 
\centering
\label{fig:islanding_diagram}
\end{figure}

\nrc{In both TM and BEM-PC approaches, instead of  network reference frame ($R-I$ frame) rotating at synchronous speed $\omega_s$, we project all phasors of an island on the COI frame ($d_{coi}-q_{coi}$)~\cite{kundur_book,sauer_pai_book} rotating at $\omega _{coi}  = \frac{1}{{H_T }}\sum\limits_{i \in M} {H{}_i} \omega _i $, where $\omega_i$ and $H_i$ are the rotor speed and inertia constant of the $i$th machine, $H_T = \sum\limits_{i \in M} {H{}_i}$, and $M$ is the set of indices indicating machine numbers in the corresponding island. Figure \ref{fig:RI_dq_coi_quad} shows the terminal voltage phasor of the $i$th machine projected on different reference frames including the machine's own  $d-q$ frame. The use of COI frame leads to rotor angle  $\theta_i = \delta_i - \delta_{coi}$ (where, $\delta _{coi}  = \frac{1}{{H_T }}\sum\limits_{i \in M} {H_i } \delta _i$), which is constant in \textit{steady state} under off-nominal frequency. This helps in efficient convergence of Netwon iterations in \eqref{eq_inverse_jac}, which is common knowledge. What is challenging however, is adapting this framework for a cascading scenario that leads to formation of multiple children from a parent island, see Fig.~\ref{fig:islanding_diagram}.}

\vspace{-4pt}

\vspace{4pt}
\nrc{\textit{Challenge during formation of multiple islands:} Let $t = t_n$ be the last instant when the parent island was intact and $t = t_{n+1}$ be the first post-islanding instant froming children as in Fig.~\ref{fig:islanding_diagram}. We need to use dynamic states $x_n$ to predict $x_{n+1}$ for both TM \eqref{eq_TM} and BEM-PC \eqref{eq_BEM}, and in addition $V_n$ is required for TM. Since $M_p \neq M_{ci} \forall i$, see Fig.~\ref{fig:islanding_diagram}, the converged $x_n$ values corresponding to the pre-islanding instant cannot be used. To be more specific, the challenge comes from the fact that unlike the $R-I$ frame which can be applied for any island, the COI frames are locally applicable to individual islands (Fig.~\ref{fig:islanding_diagram}). To solve this problem, we propose an adaptive COI frame-based approach, which is described next.}

\nrc{\textit{Proposed approach:} We perform the following steps to calculate $\left[ {\begin{array}{*{20}c}
   {x_{n + 1}^T } & {V_{n + 1}^T }  \\
\end{array}} \right]^T $ within any child island $\#ci$ --}

\nrc{\underline{Step (I):} Since $R-I$ frame is universal, we calculate $\delta_n$, $\Delta \omega_n$ $\in \mathbb{R}^{|M_{ci}|}$ as
\begin{equation}
\begin{array}{l}\label{eq_delta_COI_init}
\delta_{n} =  \theta_n + \delta_{COI\_n}^p;~~
\Delta\omega_{n} =  \Delta\bar{\omega}_n + \Delta\omega_{COI\_n}^p
\end{array}
\end{equation} where, $\theta_n ,   \Delta\bar{\omega}_n\in \mathbb{R}^{|M_{ci}|}$ are rotor angle and speed deviation vectors in island $\#ci$ w.r.t. the parent's COI frame.}

\nrc{\underline{Step (II):} This is the step where we \textit{adaptively} change the COI frames from parent to child for each island. To that end we update $\theta_n ,   \Delta\bar{\omega}_n\in \mathbb{R}^{|M_{ci}|}$ from the parent's COI frame to the COI frame of island $\#ci$ (Fig.~\ref{fig:islanding_diagram}) as 
\begin{equation}
\begin{array}{l}\label{eq:InitDelnOmega}
\theta_{n}^{ci} =  \delta_{n} - \delta_{COI\_{n}}^{ci};~~
\Delta\bar{\omega}_{n}^{ci} =  \Delta\omega_{n} - \Delta\omega_{COI\_{n}}^{ci}
\end{array}
\end{equation}
where, $\delta_{COI{\_}{n}}^{ci} = \frac{1}{H_{T,ci}}{\sum\limits_{i\in M_{ci}} H_i \delta_{in}}$ and $\Delta\omega_{COI{\_}{n}}^{ci} = \frac{1}{H_{T,ci}} {\sum\limits_{i\in M_{ci}} H_i \Delta\omega_{in}}$. Note that the other machine states, exciter states, and governor states are not changed. The device states along with $\delta_{COI\_{n}}^{ci}$ and $\Delta\omega_{COI\_{n}}^{ci}$ constitute the updated state vector $x_n\in \mathbb{R}^{6|M_{ci}|+2}$ for island $\#ci$.}

\nrc{\underline{Step (III):} We update $V_n$ within island $\#ci$ by iteratively solving \eqref{eq_mismatch_alg} for $t = t_n$ with updated states $x_n$ from Step (II). To that end, we update $Y_N$ if needed and make use of the $J_{22}$ sub-matrix of the Jacobian in \eqref{eq_jacob}.}

\nrc{\underline{Step (IV):} In the final step, we use updated $x_n$ and $V_n$ as the initial guess for $t = t_{n+1}$, i.e., we use $\left[ {\begin{array}{*{20}c}
   {\left( {x_{n + 1}^0 } \right)^T } & {\left( {V_{n + 1}^0 } \right)^T }  \\
\end{array}} \right]^T  = \left[ {\begin{array}{*{20}c}
   {\left( {x_n } \right)^T } & {\left( {V_n } \right)^T }  \\
\end{array}} \right]^T $. We iteratively solve \eqref{eq_inverse_jac} to find $\left[ {\begin{array}{*{20}c}
   {x_{n + 1}^T } & {V_{n + 1}^T }  \\
\end{array}} \right]^T $ in island $\#ci$.}

\nrc{Note that steps (II) and (III) constitute a \textit{sequential approach} within the simultaneous solution process. For $t > t_{n+1}$, the adaptive COI-frame based approach \textit{is identical with standard COI-based approach in the island's own COI frame until it further breaks into multiple islands.}}

\section{Case Studies}\label{sec:CaseStudies}

The IEEE $118$-bus system and the Polish network during winter $1999-2000$ peak condition  \cite{MATPOWER} are studied here to contrast \nrc{our proposed approach (called BEM-PC hereafter)} and the traditional approach (called TM from now). \nrcR{The IEEE $68$-bus system is also studied, which will be introduced later.} The IEEE $118$-bus system consists of $118$ buses, $54$ machines, and $186$ branches. The Polish system is a large-scale network with $2,383$ buses, $327$ machines, and $2,896$ lines. \nrc{We synthetically generate dynamic data, for these models. For the IEEE 118-bus and the Polish systems, cascades are triggered with $2$ and $3$ initial node outages, respectively, which are sufficient to create long term cascading sequences in these networks.}
For each system, $500$ Monte-Carlo runs are performed with random selection of initial node outages. \nrc{For BEM-PC we have used $\Delta t_{min} = 0.02$ s, $\Delta t_{max} = 0.4$ s, $k = 6$, $r = 7$, $\epsilon = 10^{-4}$, and maximum allowable Newton iterations is $10$.}  
\sina{For relays, $T_w^{UVLS}=3$ s, $T_{tp}^{UVLS}=3$ s, $\lambda = 25$ \%, $v_{th} = 0.8645$ pu for IEEE $118$-bus system, and $0.75$ pu for Polish system, $K^{shed}_{max}=5$, and $T_{w}^{OC}=1$ s have been used.}


\nrc{Note that the following results implement the Predictor subprocesses \textbf{(b)} of BEM-PC in Fig.~\ref{fig:fig_flow_chart_predictive_corrective} in a \textit{serial} fashion. \textit{Hence the speedup obtained is a conservative estimate of what can be obtained with parallelization of the Predictor.}  For IEEE $118$-bus system, the simulations were run in} \sina{AMD Ryzen 7 3800X CPU with $32$ GB RAM and for Polish system 4 servers with 2.2 GHz Intel Xeon Processor, 24 CPU/server, and 128 GB RAM
in PSU's ROAR facility  \cite{PSU_ROAR} were used.}

\subsection{Monte-Carlo Simulation}\label{sec:MCsimu}
\subsubsection{IEEE $118$-Bus System}

\begin{figure}[!b]
\centerline{\includegraphics[scale = 0.2, trim= 2.2cm 0cm 3.85cm 0.5cm, clip=true]{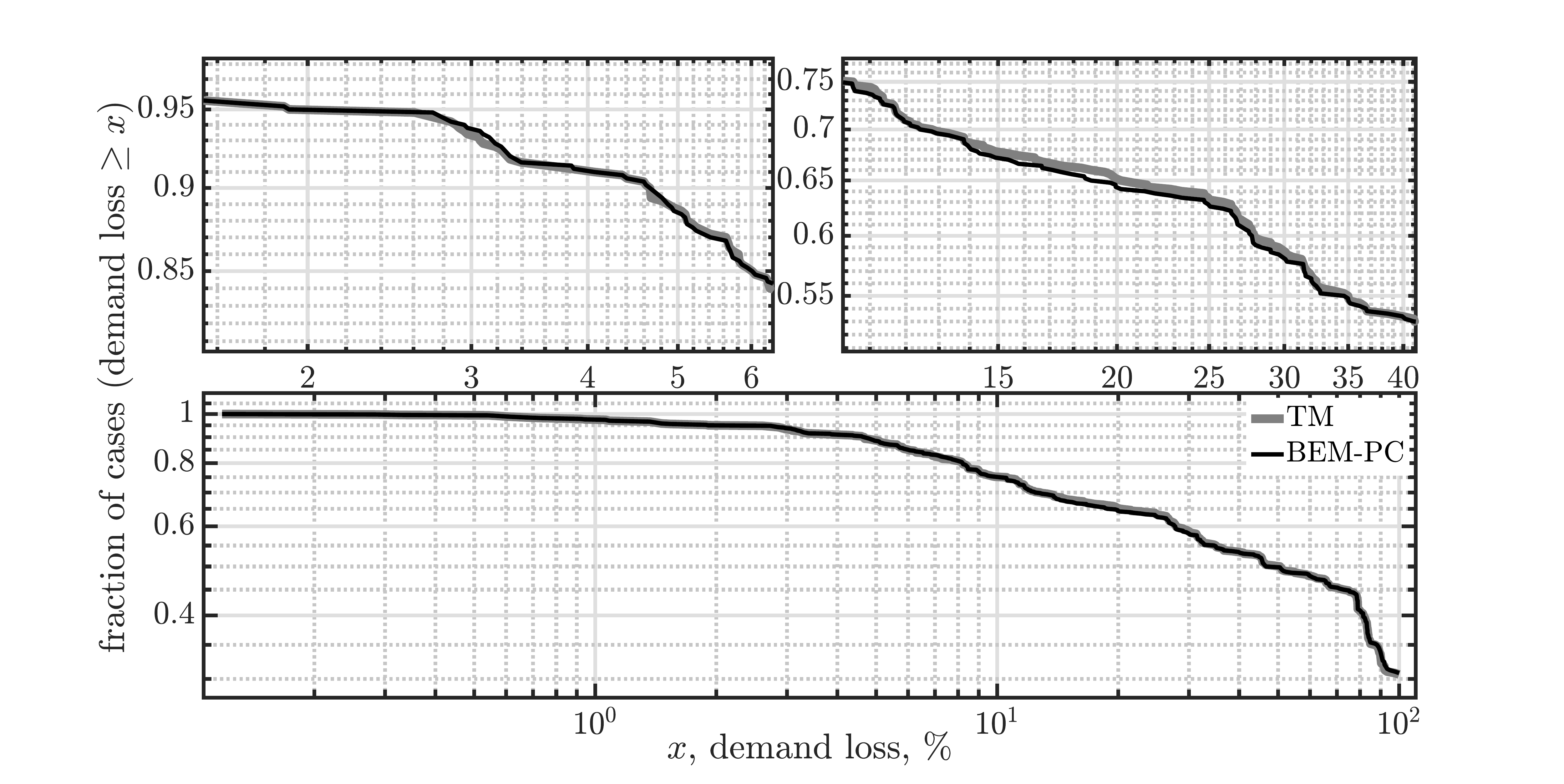}}
\caption{Fraction of cases with $\%$ demand loss $\geq x$ at the end of cascade: IEEE 118-bus system.} 
\centering
\label{distribution_demand_loss_118}
\end{figure}

\begin{figure}[!b]
\centerline{\includegraphics[scale = 0.2, trim= 2.2cm 0cm 3.85cm 0.5cm, clip=true]{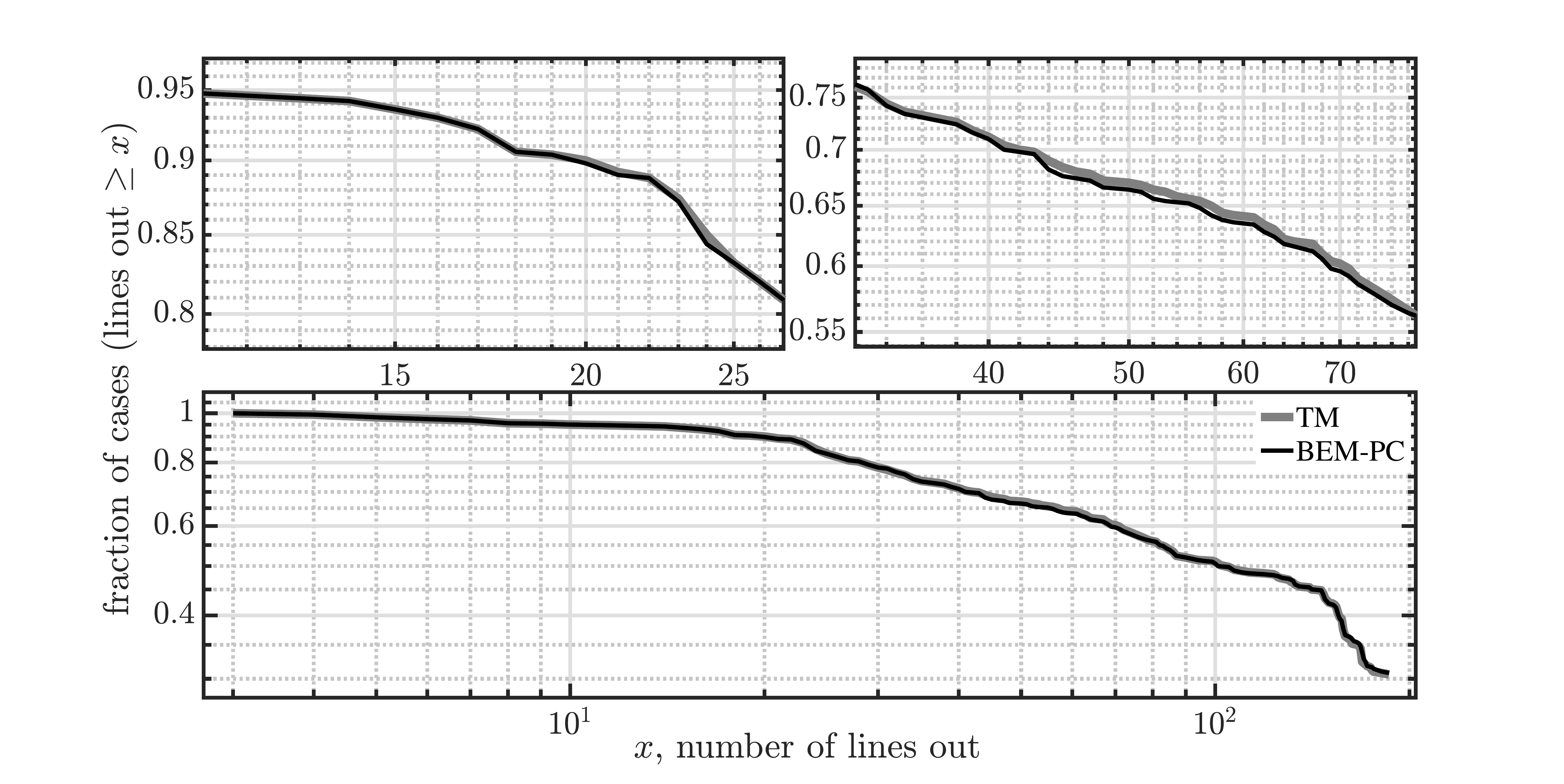}}
\caption{Fraction of cases with line outage $\geq x$ at the end of cascade: IEEE 118-bus system.} 
\centering
\label{distribution_line_outage_118}
\end{figure}

\nrc{Figures \ref{distribution_demand_loss_118} and \ref{distribution_line_outage_118} compare how often the total demand loss and line outages at the end of cascade are above a particular level for TM and BEM-PC. The analysis includes initial node outages.
The top two zoomed subplots  show that there are small differences between BEM-PC and TM for cases with $(15\%-25\%)$ demand loss and cases with $(44-60)$ line outages. Nonetheless, the results indicate a very close match between the end results of cascade. }  

\nrc{Table I compares the accuracy of BEM-PC with respect to TM and shows that  the average error at the end of cascade in states (connected vs disconnected) of buses, machines, and lines 
which are small fractions of the corresponding total numbers. Similarly, the central tendency measures of maximum errors in voltage magnitudes, angles, and frequency are very small. Although there are some outliers causing an increase in the average error values, for almost all of the cases BEM-PC is able to replicate the exact end-result of TM.}

\nrc{The $R$ values in the table show path agreement measure ~\cite{dyn2_song} between BEM-PC and TM based on dependent branch outages, where both models are subject to the same set of initial outages $C=\{c_1,c_2,...\}$. If contingency $c_i$ results in the set $A_i$ of dependent line outages in the first model and the set $B_i$ of
dependent line outages in the second model, then R is defined as follows~\cite{dyn2_song},
\begin{equation}
\begin{array}{l}\label{eq_R}
R = \frac{1}{|C|}{
\sum\limits_{i = 1}^{\left| C \right|} \frac{|A_i \cap B_i|}{|A_i \cup B_i|}}
\end{array}
\end{equation}
where, $R=1$ indicates a complete match between cascade paths from two models following all contingencies.}

\nrc{Based on the central tendency measures of $R$ from Table I and its standard deviation being $0.05916$, we conclude that the models have a high agreement in the cascade path. \textit{In addition to the high accuracy of BEM-PC in most of Monte-Carlo runs, on average it is approximately $10$ times faster than  TM.}}

 
\begin{table}[!]
\centering
\label{tab:miss_false_trip_table} 
\caption {\nrc{(a) End of cascade error, (b) path agreement measure, and (c) run time in TM w.r.t. BEM-PC: IEEE 118-Bus System}}\vspace{-0.3cm}

\begin{tabular}{cc|c|c|c|c}\\\hline\hline
                &                 & mean     & min      & max      & median   \\\hline
\multicolumn{1}{c|}{\multirow{1}{*}{error in}}         & buses           & 0.6460    & 0        & 77       & 0        \\
\multicolumn{1}{c|}{\multirow{1}{*}{state of}}         & machines        & 0.2940    & 0        & 36       & 0        \\
\multicolumn{1}{c|}{}  & lines           & 0.8960    & 0        & 114      & 0        \\\hline
\multicolumn{1}{c|}{\multirow{1}{*}{maximum}}          & $|v|, pu$         & 0.0024 & 0        & 0.0741  & 7.9$e{-7}$  \\
\multicolumn{1}{c|}{\multirow{1}{*}{error in}}         & $ \angle v, deg.$         & 0.2874 & 0        & 14.5731 & 1.0$e{-4}$ \\
\multicolumn{1}{c|}{}                & $f, Hz$            & 0.0442 & 0        & 2.3127 & 1.1$e{-6}$ \\\hline

\multicolumn{2}{c|}{R}             & 0.9911 & 0.25 & 1        & 1       \\\hline

\multicolumn{2}{c|}{\textbf{runtime ratio}} & \textbf{9.9575} & \textbf{0.3403} & \textbf{60.4361 }& \textbf{9.0554}\\\hline
\end{tabular}
\end{table}

\begin{figure}[!b]
\centerline{\includegraphics[scale = 0.2, trim= 2.2cm 0cm 3.85cm 0.5cm, clip=true]{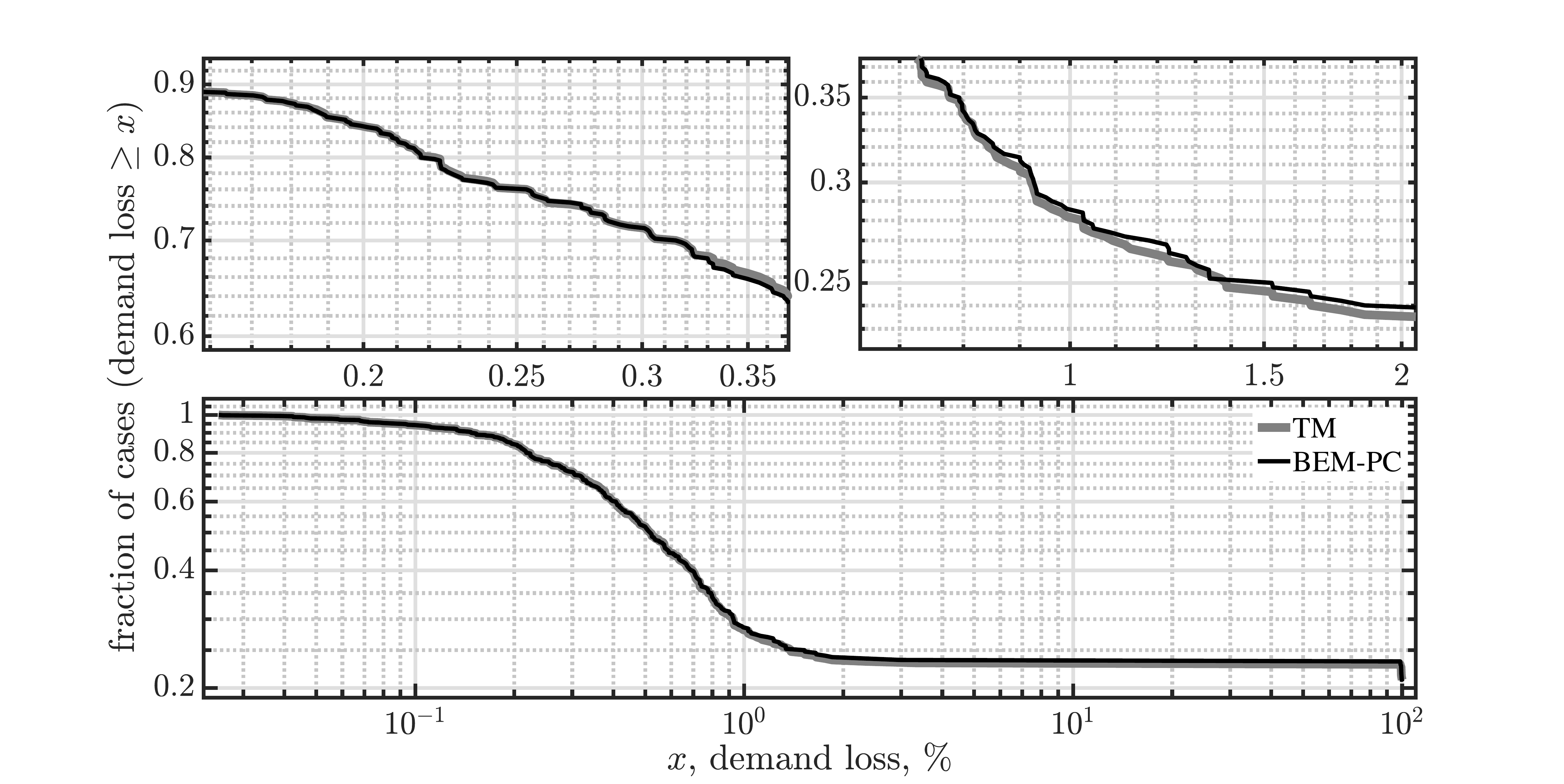}}
\caption{Fraction of cases with $\%$ demand loss $\geq x$ at the end of cascade: Polish system.} 
\centering
\label{cumulative_prob_deman_loss_Polish}
\end{figure}

\begin{figure}[!t]
\centerline{\includegraphics[scale = 0.2, trim= 2.2cm 0cm 3.85cm 0.7cm, clip=true]{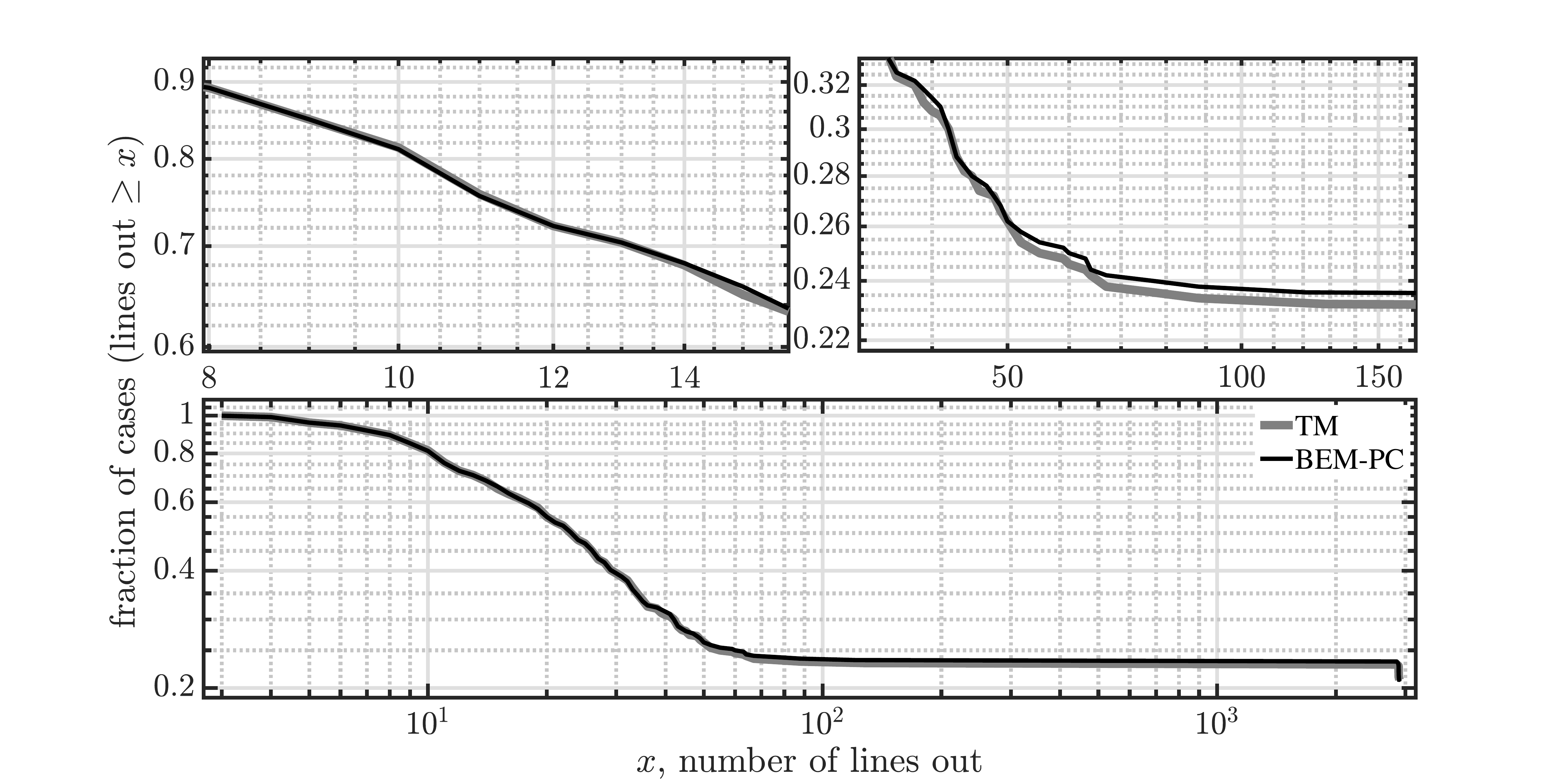}}
\caption{Fraction of cases with line outage $\geq x$ at the end of cascade:  Polish system.} 
\centering
\label{cumulative_prob_line_out_Polish}
\end{figure}

\subsubsection{Polish System}

\nrc{As before, Figs \ref{cumulative_prob_deman_loss_Polish} and \ref{cumulative_prob_line_out_Polish} compare how often the total demand loss and line outages at the end of cascade are above a particular level for TM and BEM-PC -- a very close match is observed.}

 \begin{figure*}[!h]
    \centering
    \begin{subfigure}{0.5\textwidth}
        \centerline{\includegraphics[scale = 0.19, trim= 0.0cm 0cm 0cm 0cm, clip=true]{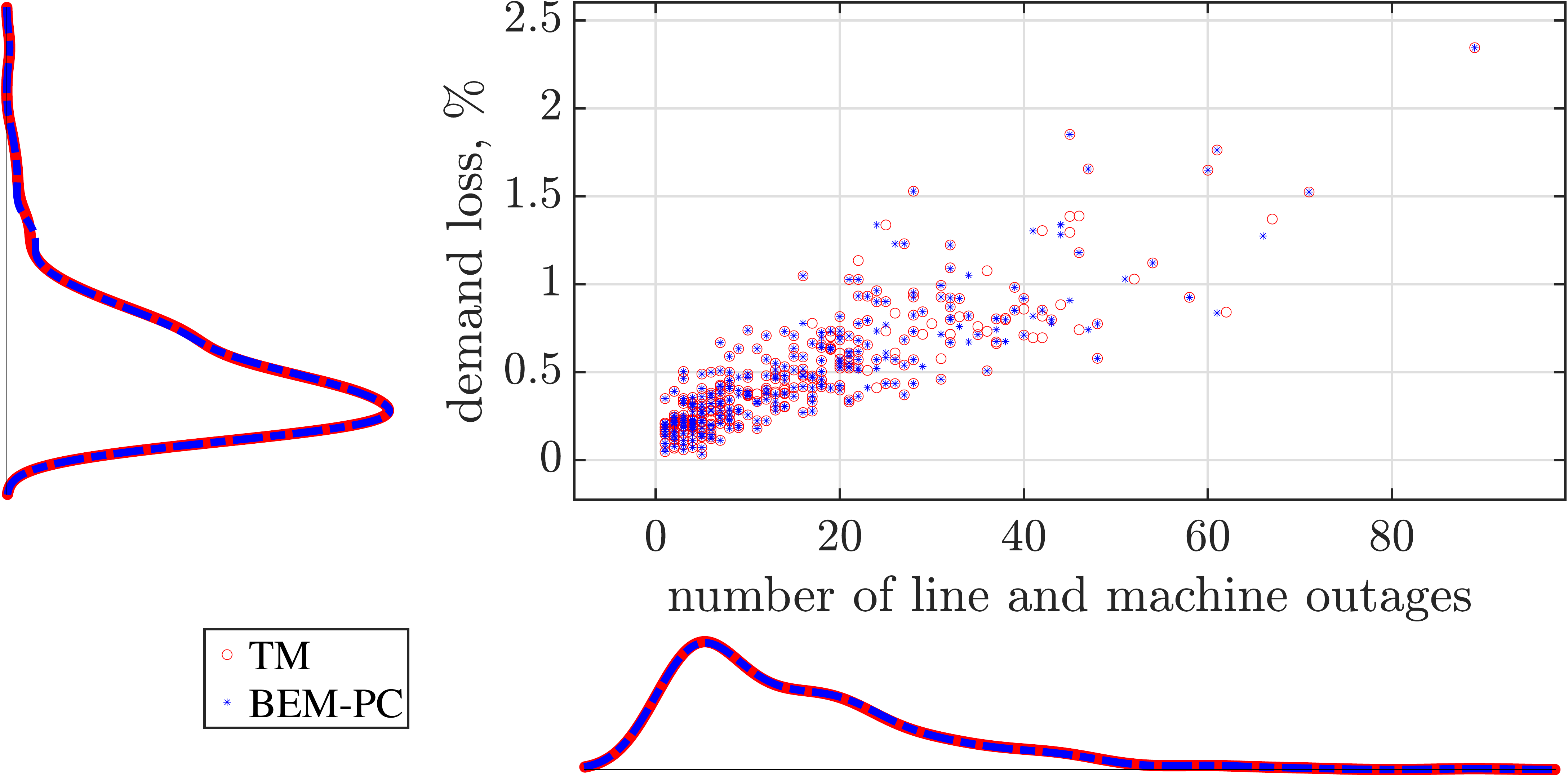}}
    \end{subfigure}%
    \begin{subfigure}{0.5\textwidth}
        \centerline{\includegraphics[scale = 0.19, trim= 0.0cm 0cm 0cm 0cm, clip=true]{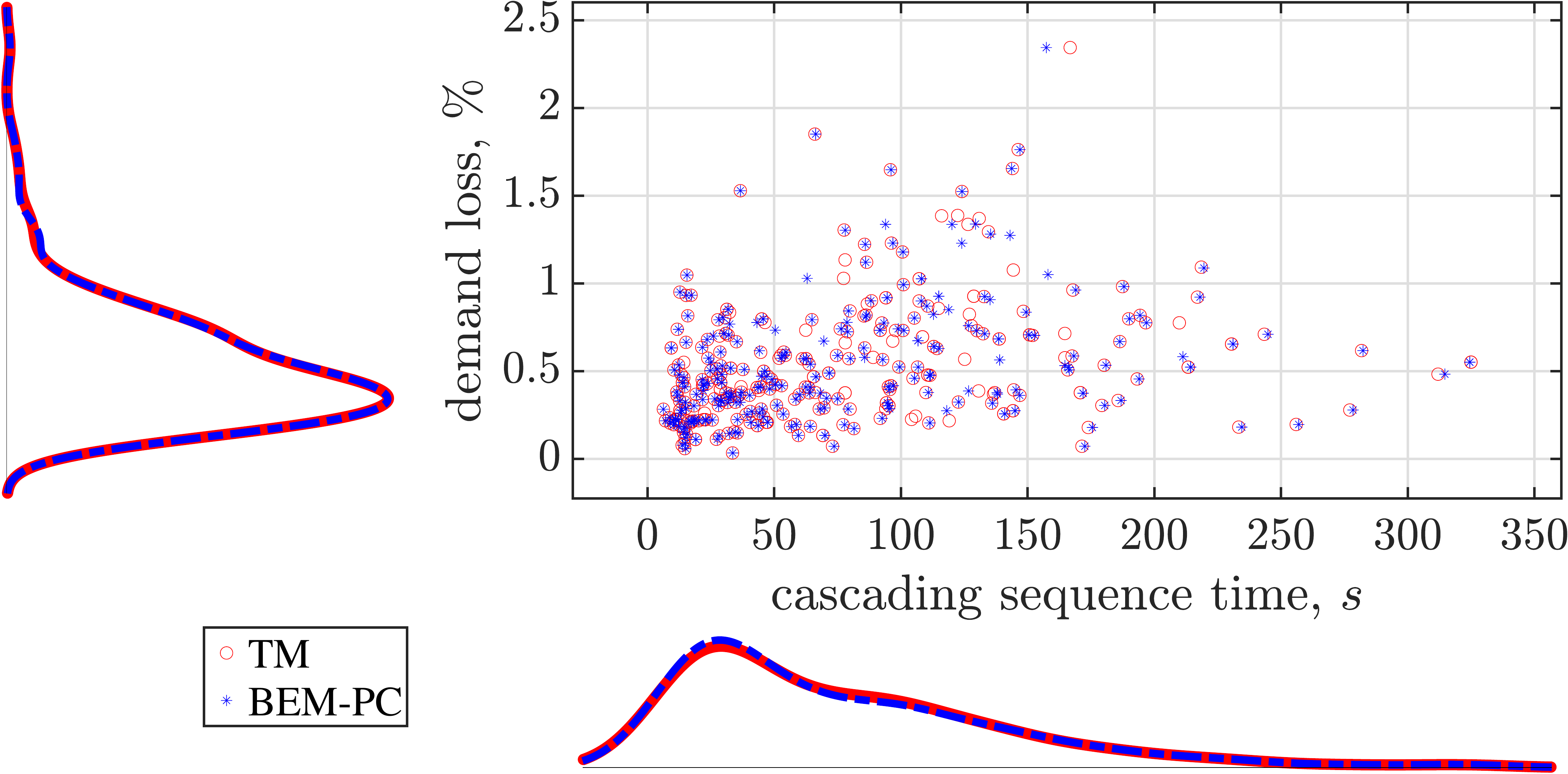}}
    \end{subfigure}
    \caption{{\color{black} Left: Demand loss vs number of branch and machine outages. Right: Demand loss vs cascade sequence time for Polish system. The density curves of the $x$ and $y$ variables are shown on the bottom and left of each subplot, respectively.}}
    \label{fig_scatter_Polish_sys}
\end{figure*}



 \begin{table}[!t]
\centering
\label{tab:miss_false_trip_table} 
\caption {\nrc{(a) End of cascade error, (b) path agreement measure, and (c) run time in TM w.r.t. BEM-PC: Polish System}}\vspace{-0.3cm}

\begin{tabular}{cc|c|c|c|c}\\\hline\hline
                &                 & mean     & min      & max      & median   \\\hline
\multicolumn{1}{c|}{\multirow{1}{*}{error in}}         & buses           & 0.1220    & 0        & 8       & 0        \\
\multicolumn{1}{c|}{\multirow{1}{*}{state of}}         & machines        & 0.0620    & 0        & 3       & 0        \\
\multicolumn{1}{c|}{}  & lines           & 0.1600    & 0        & 7      & 0        \\\hline
\multicolumn{1}{c|}{\multirow{1}{*}{maximum}}          & $|v|, pu$         & 0.0008   & 0        & 0.0360  & 2.1$e{-5}$  \\
\multicolumn{1}{c|}{\multirow{1}{*}{error in}}         & $ \angle v, deg.$         & 0.1441 & 0        & 10.1075 & 4.0$e{-4}$ \\
\multicolumn{1}{c|}{}                & $f, Hz$            & 0.0165 & 0        & 0.2414 & 8.4$e{-5}$ \\\hline
\multicolumn{2}{c|}{R}             & 0.9922
 & 0.75
 & 1        & 1       \\\hline

\multicolumn{2}{c|}{\textbf{runtime ratio}} & \textbf{34.6097} & \textbf{1.1593} & \textbf{430.3984} & \textbf{24.7959}\\\hline
\end{tabular}
\end{table}



\nrc{Table II compares various error measures at the end of cascade for BEM-PC with respect to TM. These indicate that for almost all of the cases, BEM-PC is able to accurately mimic the end results of cascade as in TM. The central tendency measures of $R$ from Table II and its standard deviation equalling $0.03230$ demonstrate that the models have a high degree of agreement in the cascade path. \textit{Finally, $runtime~ratio$ indicates that on average the proposed model is $34.6$ times faster than the standard model. }}


\nrc{Figure  \ref{fig_scatter_Polish_sys} provides comparison between BEM-PC and TM on correlation of various end-of-cascade measures like demand loss  vs number of outages and demand loss vs cascade sequence time (time between initial and final events). Out of $500$ Monte-Carlo runs with $3$ initial node outages, $41$ cases are resilient cases and did not lead to any dependent events after initial node outages, whereas $118$ cases did not converge and were considered as collapsed cases. In both panels in this figure, cases without dependent events and cases with complete collapse are disregarded. Following are the observations}
\nrc{\begin{itemize}
    \item Both plots indicate that the correlations among the two pairs of variables in BEM-PC closely match that of TM. However, there are a few cases in which these two approaches produce slightly different results.
    \item The density curves reveal almost identical distribution patterns for TM and BEM-PC for both sets of values on $x$ and $y$ axes.
    \item The right panel indicates a considerable number of cases have cascade sequence time more than $100$ s. 
\end{itemize}}
 
 
\begin{figure}[!t]
\centerline{\includegraphics[scale = 0.2, trim= 2.2cm 0cm 3.85cm 0.2cm, clip=true]{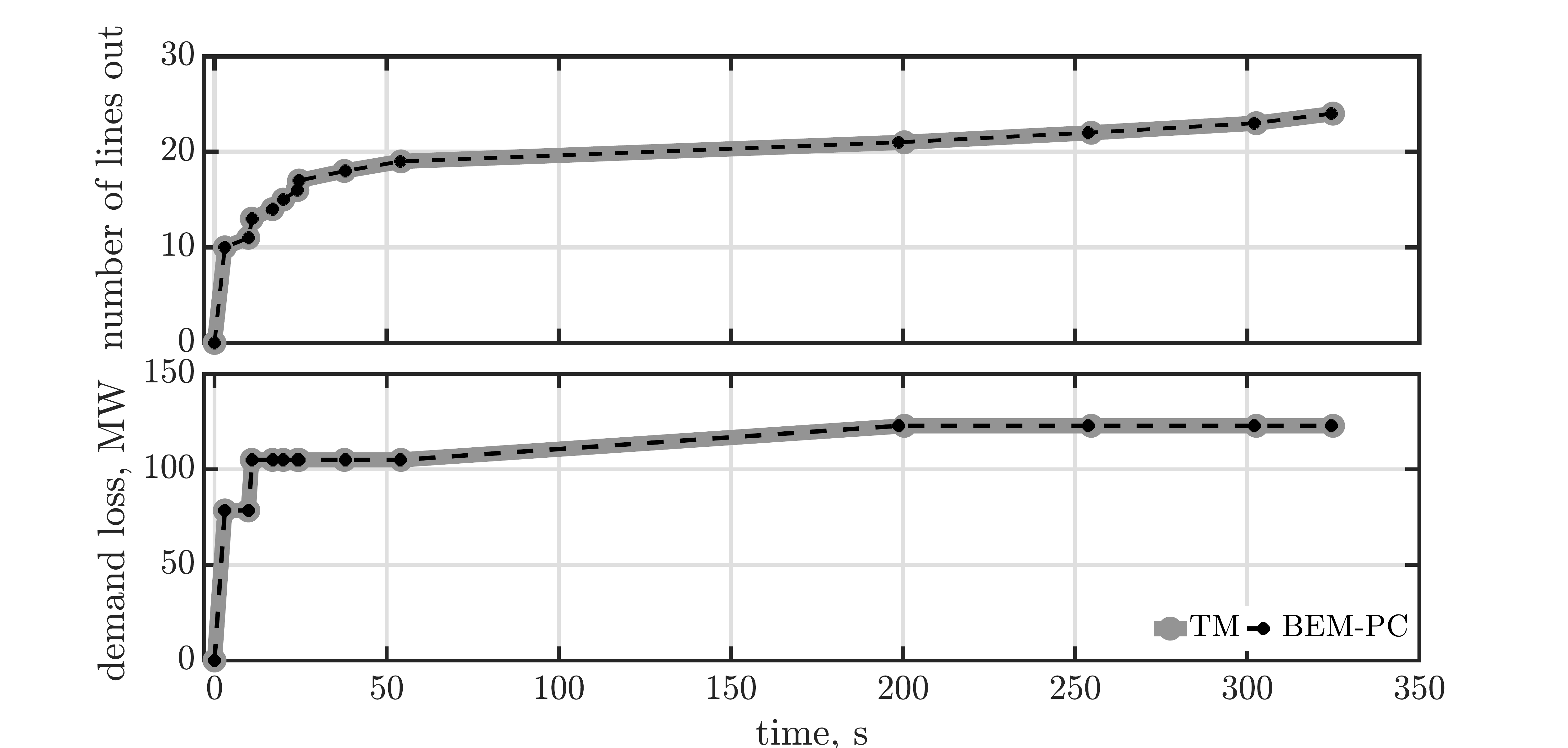}}
\caption{Timeline of number of branch outages and demand loss during cascade: Polish system.} 
\centering
\label{demand_line_time_cosmic_Polish_regular_case}
\end{figure}
\vspace{-0.2cm}

For a sample case in the Polish system, time-domain plots representing \nrc{the} number of branch outages and demand loss against time are shown in Fig. \ref{demand_line_time_cosmic_Polish_regular_case}. \nrc{The nodes in the figure indicate the instants of outage/demand loss. The plots reveal that BEM-PC is following the exact cascade path as in the ground truth. Also, it is worth noting that the proposed simulation approach is $28$ times faster than TM in this case.} \nrcR{As described earlier, initially (at $t= 3$ s) $3$ random nodes are disconnected to trigger cascade. These initial node outages are therefore not dependent outages, i.e., they do not represent the severity of cascade. In Fig. \ref{demand_line_time_cosmic_Polish_regular_case}, the initial node outage caused disconnection of $10$ lines at $t = 3$ s in addition to significant demand loss.}
\vspace{-4pt}
\nrc{\subsection{Hyperstability -- A Challenge for BEM and Performance of Predictor-Corrector Approach}\label{sec:HyperstabStudy}}

\nrc{In this section, accuracy of our proposed BEM-PC approach is tested against different cases with oscillatory instability in both IEEE $118$-bus system and the Polish network. Since the 118-bus and the Polish system does not exhibit oscillatory instability under nominal setting, we create two oscillatory instability situations in the system by making damping coefficient negative in some machines, first at the beginning of cascade (case \#1), and in a separate case somewhere in the middle of cascade (case \#2). }

\nrc{The SPS action is designed to trip two unstable machines with the highest amplitudes of oscillations upon detection of oscillatory instability. In TM this takes place through \textit{explicit} SPS action after 7.5 s and 4.5 s, respectively, in IEEE 118-bus and Polish system. However, in BEM-PC, a \textit{functional} implementation is achieved by identifying the unstable mode and participating machines using eigendecomposition of the $A$ matrix for post-event equilibrium as shown in Fig.~\ref{fig:fig_flow_chart_predictive_corrective}. Then, the suitable predetermined protection action is taken by SPS. Note that other type of SPS actions can also be taken like tripping certain lines to disconnect areas oscillating against each other.}

\subsubsection{IEEE $118$-Bus System}
\begin{figure}[!t]
\centerline{\includegraphics[scale = 0.2, trim= 2.2cm 0cm 3.85cm 0.5cm, clip=true]{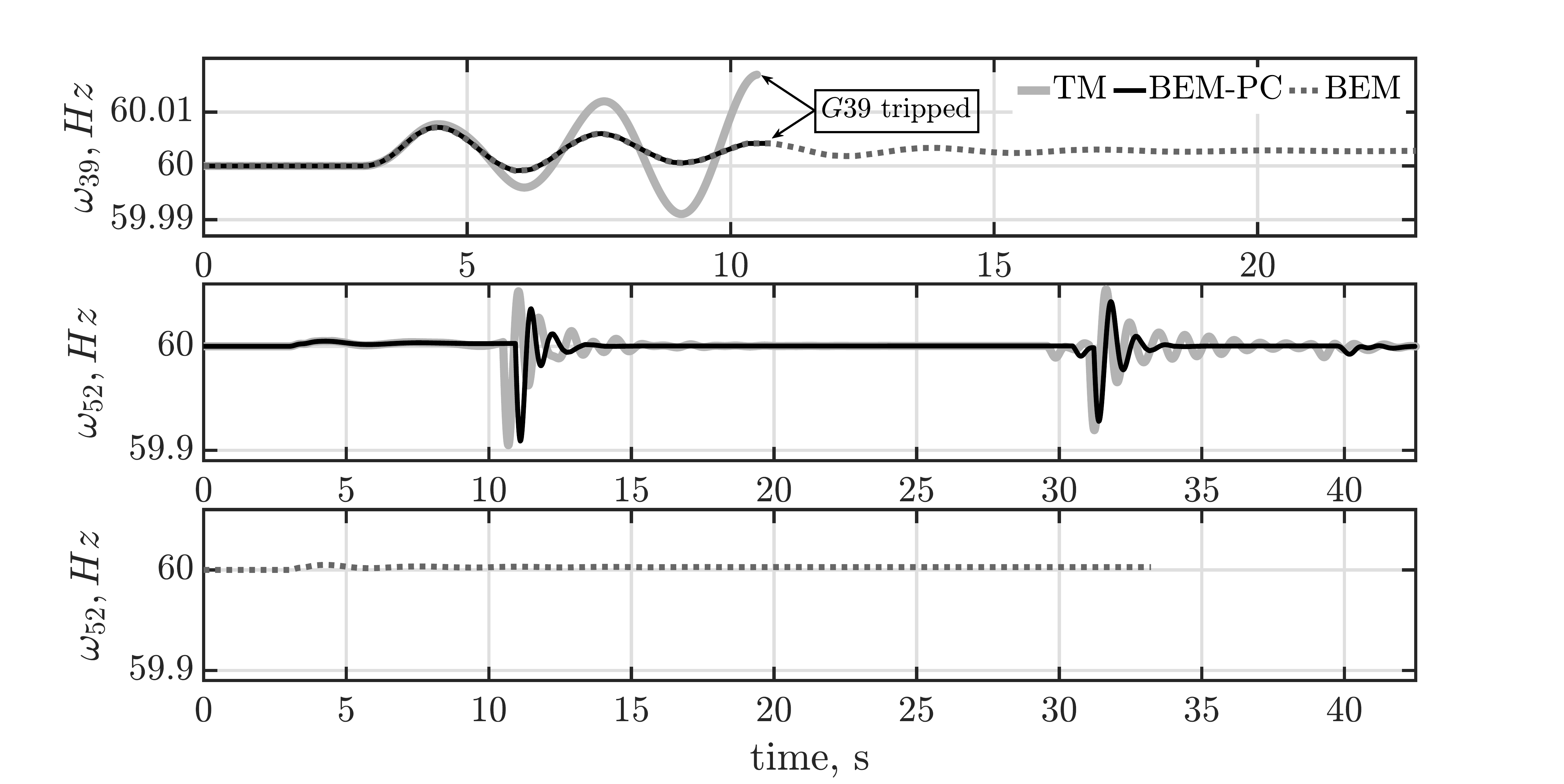}}
\caption{IEEE 118-bus system case $\#1$: Variation of speeds of $G39$ and $G52$ with oscillatory instability from beginning of cascade.} 
\centering
\label{figure_osc_first_IEEE_final}
\end{figure}

\begin{figure}[!t]
    \centering
    \begin{subfigure}{0.5\textwidth}
        \centerline{\includegraphics[scale = 0.16, trim= 0.0cm 0cm 0cm 0.0cm, clip=true]{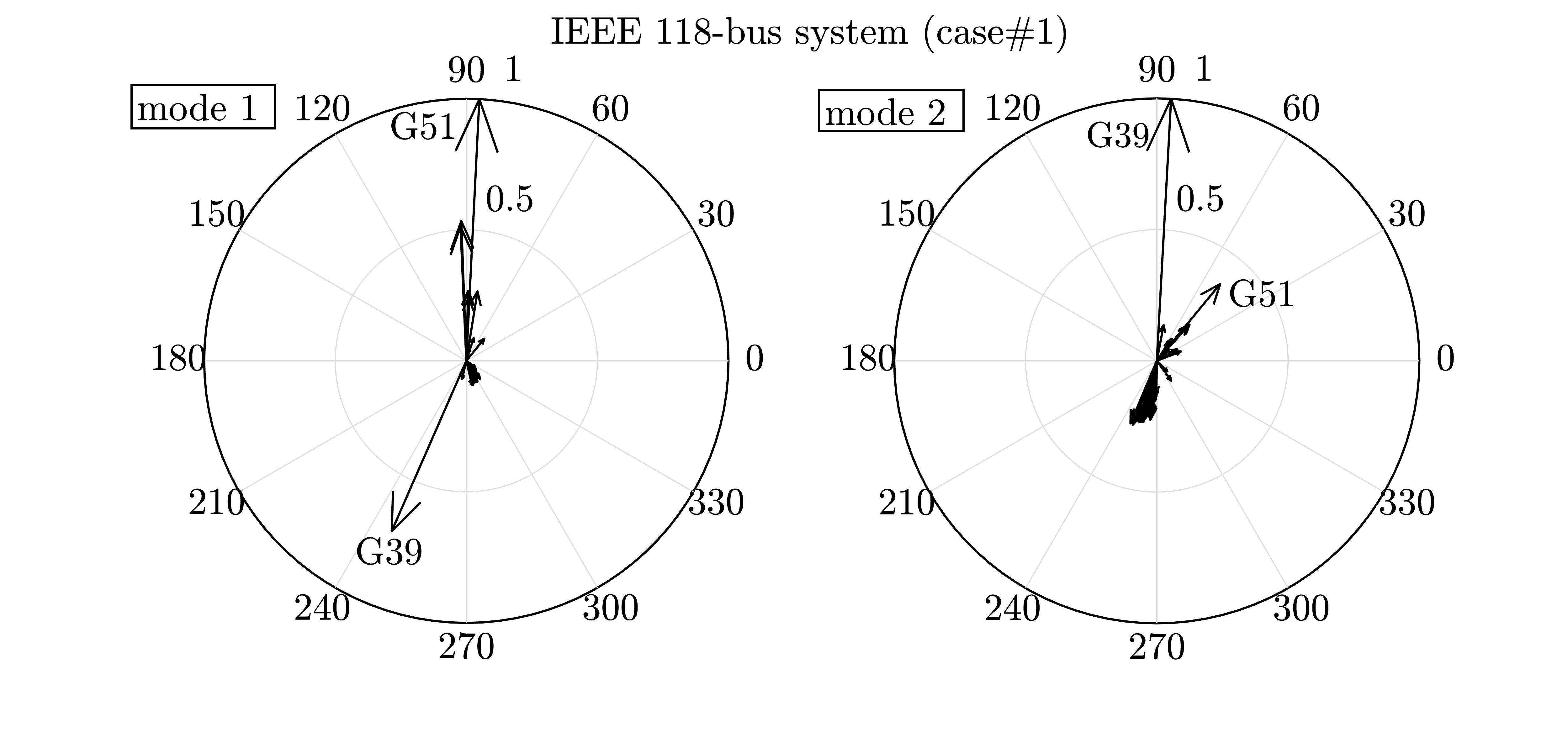}}
    \end{subfigure}\\ \vspace{-12pt}
    \begin{subfigure}{0.5\textwidth}
        \centerline{\includegraphics[scale = 0.16, trim= 0.0cm 0cm 0cm 0.0cm, clip=true]{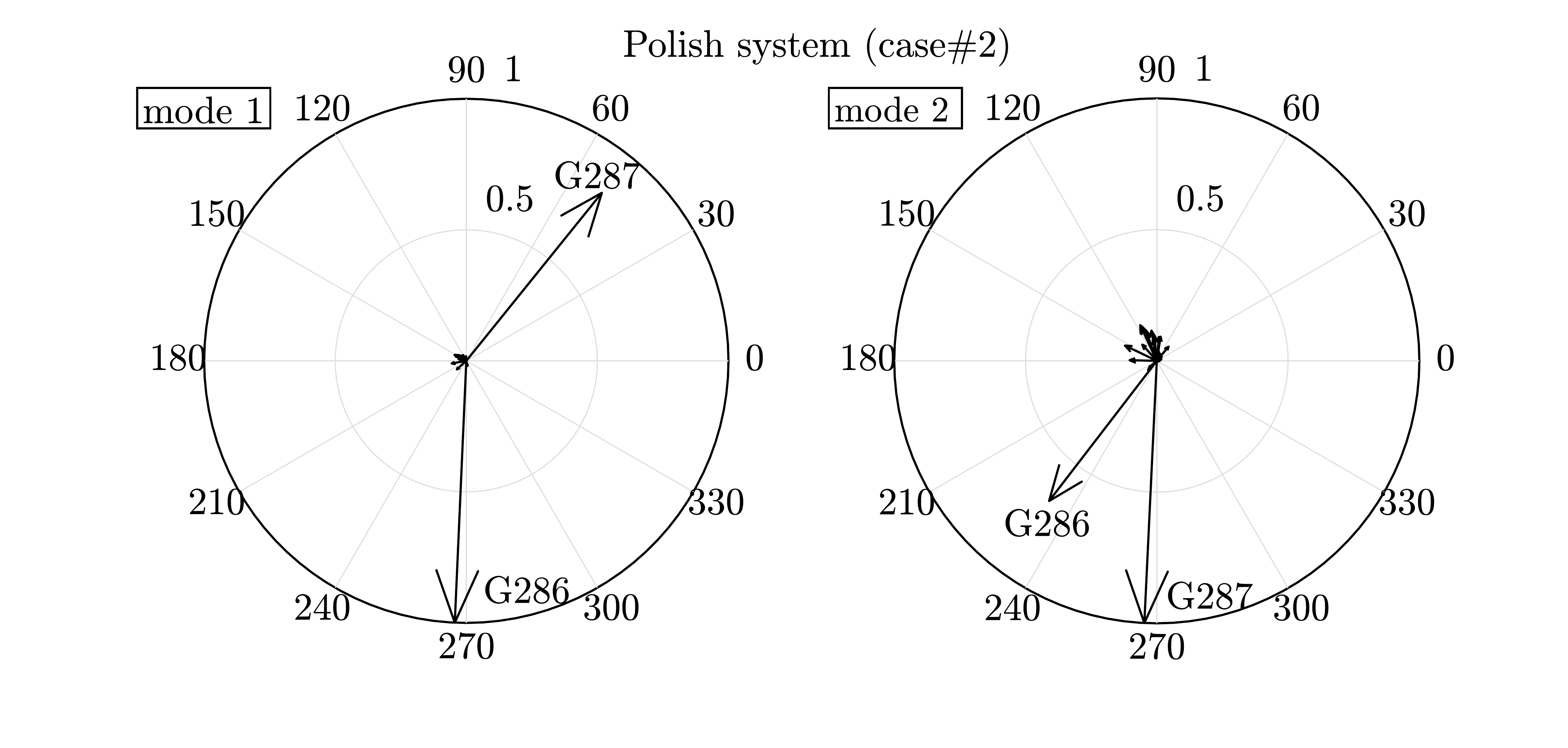}}
    \end{subfigure}
    \caption{\nrc{Modeshapes of generator speeds estimated by BEM-PC corresponding to the unstable modes for case \#1 in IEEE $118$-bus and for case \#2 in Polish system under the predictor subprocess (b) in Fig.~\ref{fig:fig_flow_chart_predictive_corrective}.} }
    \label{fig:compass}
\end{figure}

\nrc{We make the damping coefficients of generators $G39$ and $G51$ negative. For case $\#2$, we introduce the negative values at the start of third tier in the middle of cascade. Figure \ref{figure_osc_first_IEEE_final} shows rotor speeds of two representative machines in case $\#1$ for TM, BEM-PC, and BEM without PC-approach (BEM). The top and bottom subplots show that BEM leads to only one tier of cascade  due to hyperstability issue. The top panels show that BEM-PC has captured the oscillatory instability in the system and is able to tackle the hyperstability issue of BEM -- note slight difference in tripping times in BEM-PC due to the OC relays' window-based averaging described in Section~\ref{sec:COI_frame_models}. The estimated unstable modes by BEM-PC are $0.1239 \pm j2.4889$ and $0.1121 \pm j2.0531$ and the corresponding estimated modeshapes are shown in Fig.~\ref{fig:compass}. Based on the modeshapes, $G51$ and $G39$ are tripped after a $7.5$ s delay. Eventually, TM and BEM-PC lead to $25$ tiers of cascade (not shown here). }

\nrc{ Tables III and IV compare path agreement and various end-of-cascade measures for these three models in cases $\#1$ and $\#2$. Clearly, BEM-PC solved the hyperstability issue of BEM, had identical cascade propagation path, and replicated the end results of cascade in the ground truth in much shorter time. As expected, BEM without PC approach  shows significantly different results than the ground truth.}

\begin{table}[!b]
\label{tab:miss_false_trip_table} 
\centering
\caption {End-of-cascade comparison: IEEE $118$-Bus System}
\resizebox{0.95 \columnwidth}{!}{
\begin{tabular}{c|c|c|c|c|c|c}\hline\hline
\multicolumn{2}{c|}{}                   & demand  & lines & mach.  & cascade & runtime   \\
\multicolumn{2}{c|}{}                   & loss, \%  & out & out & time, s &  ratio  \\ \hline
\multirow{3}{*}{\rotatebox[origin=c]{0}{case \#1}} & TM         & 83.41     & 157       & 44          & 146.93  & 7.72 \\
                                                  & BEM-PC       & 83.41     & 157       & 44          & 146.23  & 1 \\
                                                  & BEM          & 1.06      & 4         & 0           & 3       & 0.09   \\\hline
\multirow{3}{*}{\rotatebox[origin=c]{0}{case \#2}} & TM         & 7.01      & 31        & 6           & 337.11  & 5.29 \\
                                                  & BEM-PC       & 7.01      & 31        & 6           & 336.78  & 1 \\
                                                  & BEM          & 1.88      & 9         & 0           & 69.88   & 0.27 \\\hline 
\end{tabular}}
\end{table}


\begin{table}[!h]
\label{tab:miss_false_trip_table} 
\caption {End-of-cascade and path agreement comparison: IEEE $118$-bus System}
\resizebox{1 \columnwidth}{!}{
\begin{tabular}{c|c|c|c|c|c|c|c|c}\hline\hline
\multicolumn{2}{c|}{\multirow{2}{*}{}} & \multicolumn{3}{c|}{error   in state of} & \multicolumn{3}{c|}{max   error in} & \multirow{2}{*}{R} \\\cline{1-2}  \cline{3-5} \cline{6-8}  
Case                 & TM vs                         & buses & mach. & lines & $|v|, pu$  & $\angle v, deg.$  & $f, Hz$     &          \\\hline
\multirow{2}{*}{\rotatebox[origin=c]{0}{ \#1}} & BEM-PC     & 0     & 0        & 0     & 7.8$e{-7}$ & 2$e{-4}$ & 1.8$e{-4}$ & 1        \\
                                               & BEM & 93    & 44              & 153   & 9.7$e{-2}$ & 13.58 & 4.09 & 0  \\\hline
\multirow{2}{*}{\rotatebox[origin=c]{0}{ \#2}} & BEM-PC     & 0     & 0        & 0     & 5.8$e{-6}$ & 6.4$e{-3}$ & 5.3$e{-5}$ & 1        \\
                                               & BEM & 6     & 6               & 22    & 9.7$e{-2}$ & 15.63 & 1.4$e{-4}$ & 0.19\\\hline
\end{tabular}}
\end{table}

 
 
\begin{table}[!h]
\label{tab:miss_false_trip_table} 
\caption {End-of-cascade comparison:  Polish System}
\centering
\resizebox{0.95 \columnwidth}{!}{
\begin{tabular}{c|c|c|c|c|c|c}\hline\hline
\multicolumn{2}{c|}{}                   & demand  & lines & mach.  & cascade & runtime   \\
\multicolumn{2}{c|}{}                   & loss, \%  & out & out & time, s &  ratio  \\ \hline
\multirow{3}{*}{\rotatebox[origin=c]{0}{case \#1}} & TM     & 0.96 & 34 & 4 & 90.84  & 25.74  \\
                            & BEM-PC                         & 0.96 & 34 & 4 & 90.74  & 1   \\
                            & BEM                            & 0.22 & 9  & 1 & 11.07  & 0.15    \\\hline
\multirow{3}{*}{\rotatebox[origin=c]{0}{case \#2}} & TM     & 0.92 & 35 & 3 & 115.66 & 40.37 \\
                            & BEM-PC                         & 0.92 & 35 & 3 & 115.96 & 1   \\
                            & BEM                            & 0.18 & 10 & 0 & 43.69  & 0.16   \\\hline
\end{tabular}}
\end{table}


\begin{table}[!h]
\label{tab:miss_false_trip_table} 
\caption {{End-of-cascade and path agreement comparison: Polish System}}
\resizebox{1 \columnwidth}{!}{
\begin{tabular}{c|c|c|c|c|c|c|c|c}\hline\hline
\multicolumn{2}{c|}{}          & \multicolumn{3}{c|}{error   in state of} & \multicolumn{3}{c|}{max   error in}                                                      & \multirow{2}{*}{R} \\\cline{1-2} \cline{3-5} \cline{6-8} 
Case                 & TM vs  & buses       & mach.      & lines      & $|v|, pu$                      & $\angle v, deg.$                    & $f, Hz$                        &                    \\\hline
\multirow{2}{*}{\#1} & BEM-PC & 0           & 0            & 0          & 3.2$e{-5}$                     & 1.2$e{-3}$               & 7.3$e{-5}$ & 1                  \\
                     & BEM    & 15          & 3            & 25         & 5.7$e{-2}$                     & 4.15                     & 9.7$e{-4}$ & 0.11             \\\hline
\multirow{2}{*}{\#2} & BEM-PC & 0           & 0            & 0          & 3.3$e{-5}$                     & 1.2$e{-3}$               & 7.6$e{-5}$                     & 1                  \\
                     & BEM    & 15          & 3            & 25         & 5.7$e{-2}$                     & 4.19                     & 1.1$e{-3}$                     & 0.14             \\ \hline
\end{tabular}}
\end{table}
\subsubsection{Polish System}
\begin{figure}[!t]

\centerline{\includegraphics[scale = 0.2, trim= 1.0cm 0cm 3.85cm 0.5cm, clip=true]{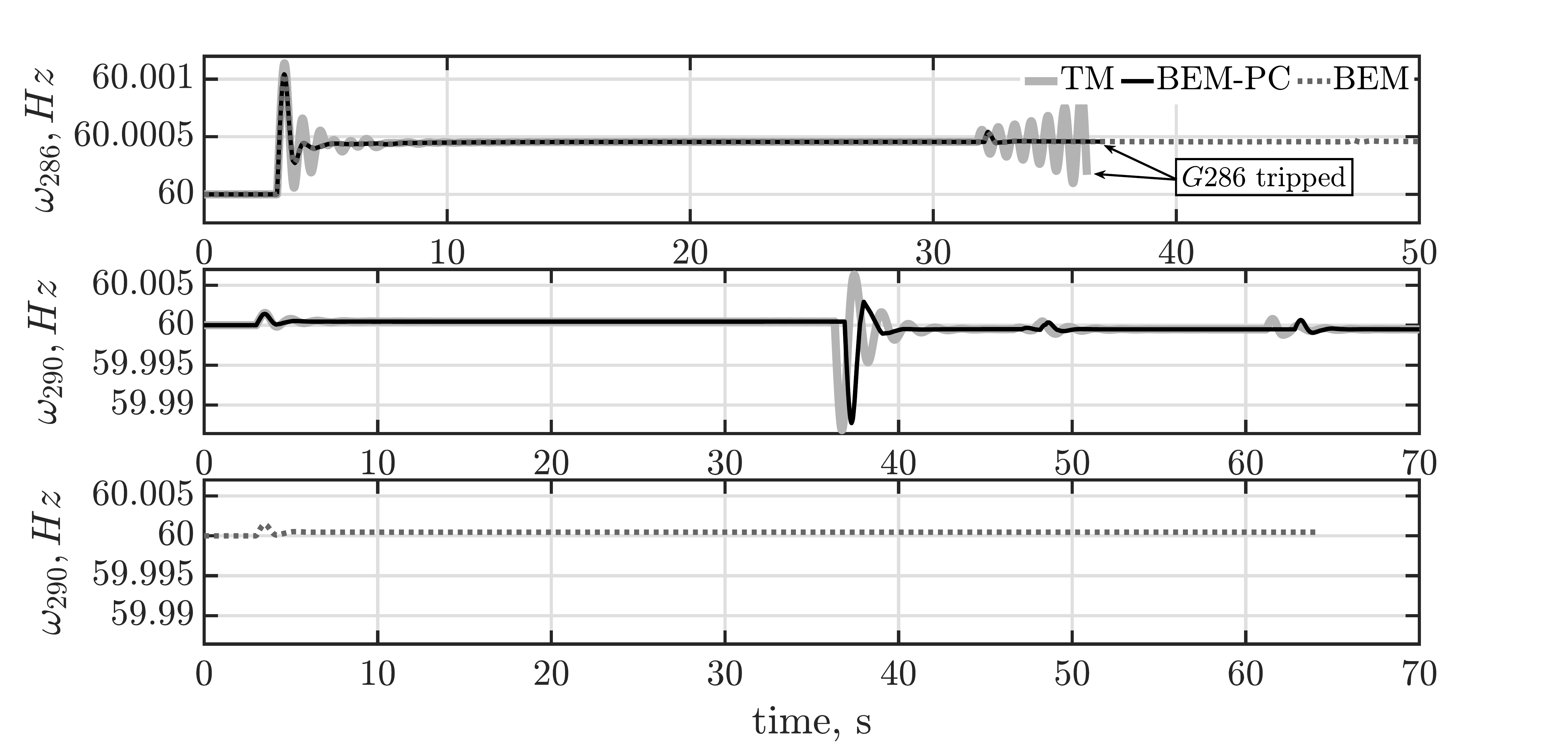}}
\caption{Polish system case $\#2$: Variation of speeds of $G286$ and $G290$ with oscillatory instability in the middle of cascade.} 
\centering
\label{OSC_instab_Polish_plot_mach}
\end{figure}

\begin{figure}[!t]
\centerline{\includegraphics[scale = 0.2, trim= 2.2cm 0cm 3.85cm 0.5cm, clip=true]{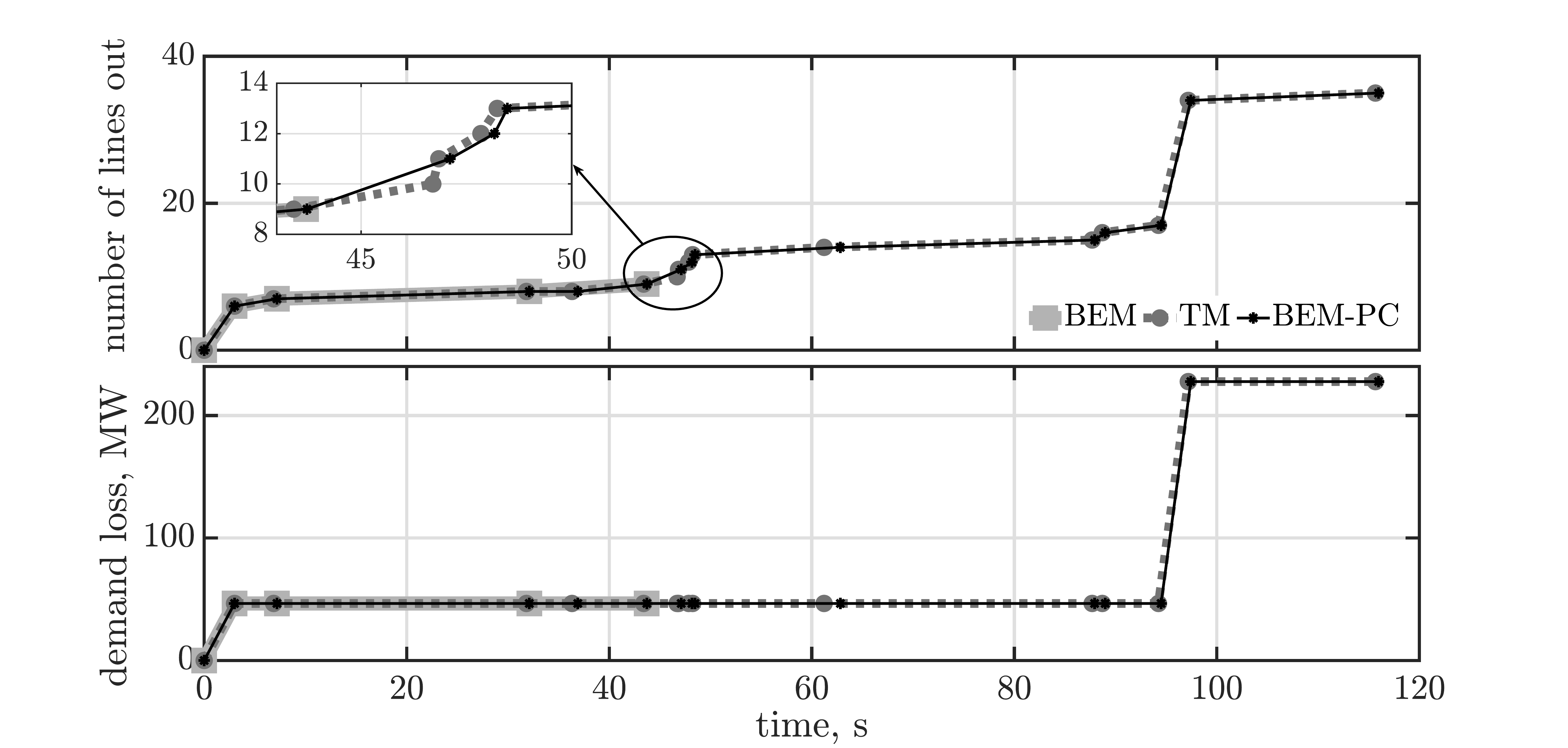}}
\caption{Timeline of number of branch outages and demand loss during cascade in the oscillatory instability case \#2: Polish system.} \centering
\label{demand_line_time_cosmic_Polish_Osc_insta}
\end{figure}

\nrc{Generators $G286$ and $G287$ are selected to create the oscillatory instability in the system. Figure \ref{OSC_instab_Polish_plot_mach} shows rotor speed variation in two selected machines in the system for case $\#2$. While BEM without PC is not able to capture the oscillatory instability and diverges from the ground truth, this figure along with Tables V and VI reveal that in both cases, BEM-PC attain the identical end results of cascade as TM. As an example, for case \#2, BEM-PC estimates the unstable modes $0.4306 \pm j9.7845$ and $0.4464 \pm j9.3910$, and corresponding modeshapes in Fig.~\ref{fig:compass}, which leads to trippings of $G286$ and $G287$ after a $4.5$ s delay by the SPS. Finally, Fig. \ref{demand_line_time_cosmic_Polish_Osc_insta} represents branch outages and demand loss against the cascade progression time for case $\#2$. It reveals that cascade in BEM is stopping after $4$ tiers around $43$ s. Although, because of very close trip delays of two overloaded lines around $t=45$ s, these lines are tripped together in one tier of cascade in BEM-PC, and in two tiers in TM, they follow  identical cascade propagation paths (see, R in Table VI) and produce the same results at the end-point of cascade.}

\begin{figure}[!t]
\centerline{\includegraphics[scale = 0.205, trim= 2.2cm 0cm 3.85cm 0.5cm, clip=true]{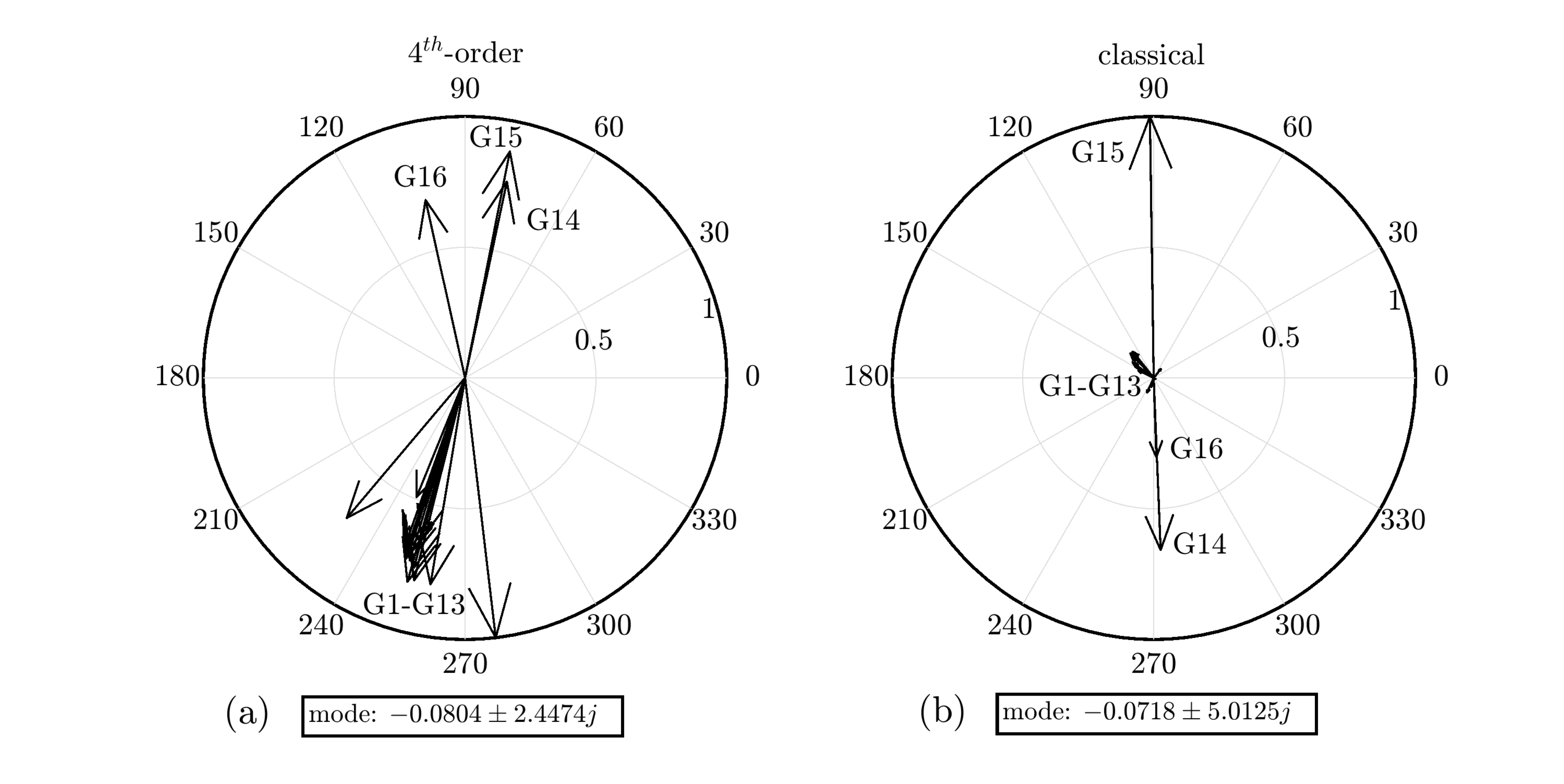}}
\caption{\nrcR{Modeshapes of generator speeds corresponding to the most poorly-damped mode for $4^{th}$-order, and classical models of NE-NY system under the predisturbance condition.}}
\centering
\label{fig:mode_shapes_predis_NENY}
\end{figure}

\nrcR{For further judging the efficacy of the proposed BEM-PC approach in handling the hyperstability issue, it would be ideal to study a system that naturally exhibits oscillatory instability as the cascade propagates. To that end, we have also considered the IEEE $68$-bus New England-New York (NE-NY) benchmark test system \cite{nrc_thesis} that is widely used for studying oscillatory instability problems. 
\subsubsection{IEEE $68$-Bus NE-NY System}\label{sec:NENyhyperstab}
The system has $5$ areas, $87$ lines, and $16$ generators --  a detailed description can be found in \cite{nrc_thesis}. The grid exhibits multiple oscillatory modes among which the mode with eigenvalues $-0.0804 \pm 2.4474j$ has the least damping ratio. The modeshapes of generator speeds for this mode is shown in Fig.~\ref{fig:mode_shapes_predis_NENY}(a). We run $500$ Monte Carlo (MC) simulations with two random initial line outages in AMD Ryzen 7 3800X CPU with $32$ GB RAM. Unlike the IEEE $118$-bus and Polish systems, we use $T_w^{OC} = 4$ s, since this system exhibits low-frequency interarea modes.}

\begin{figure}[h!]
    \centering
    \begin{subfigure}[]{0.5\columnwidth}
        \centerline{\includegraphics[scale = 0.20, trim= 15cm 0cm 15cm 0.5cm, clip=true]{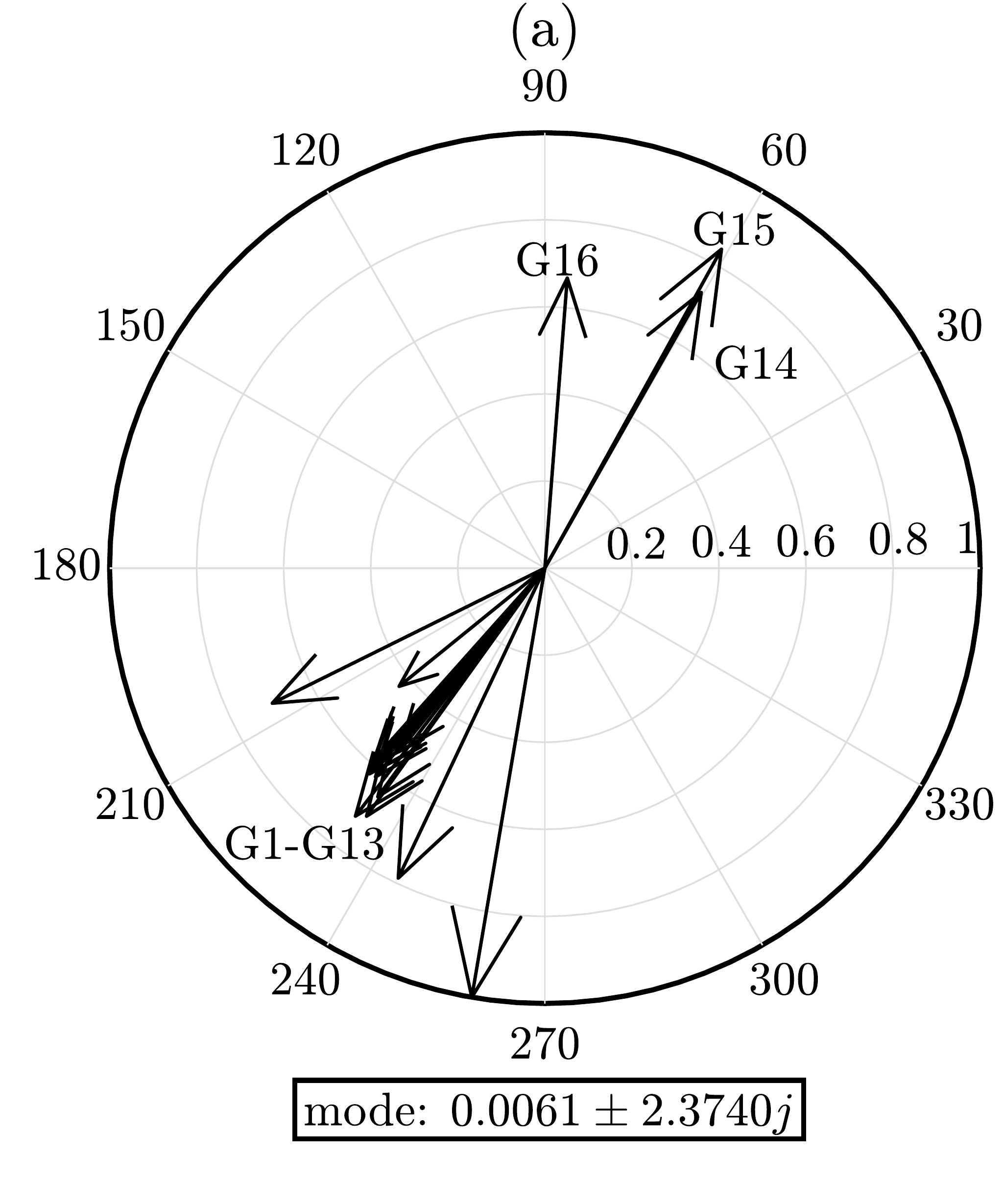}}
        \centering
    \end{subfigure}%
    \begin{subfigure}[]{0.5\columnwidth}
        \centerline{\includegraphics[scale = 0.75, trim= 0.5cm 0.5cm 0.5cm 0.5cm, clip=true]{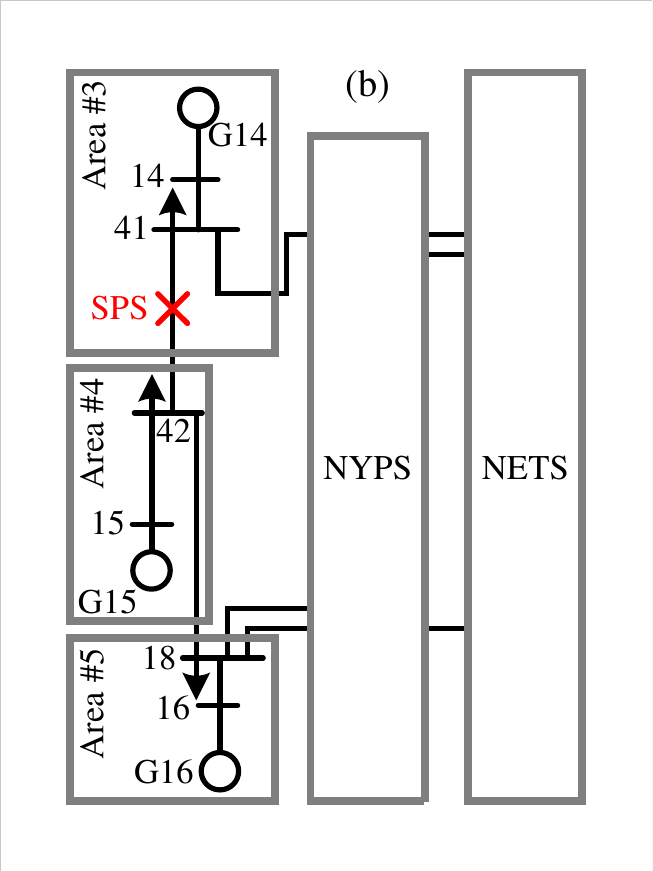}}
        \centering
    \end{subfigure}
    \caption{\nrcR{Left: Modeshapes of generator speeds estimated by BEM-PC corresponding to the unstable interarea mode for a typical case with hyperstability issue in NE-NY system under the predictor subprocess (b) in Fig.~\ref{fig:fig_flow_chart_predictive_corrective}. Right: One-line diagram of NE-NY test system highlighting SPS action.}}
     \label{fig:mode_shapes_unstable_NENY}
\end{figure}

\nrcR{As shown in Table VII, BEM-PC encounters hyperstability issues in $16.4$\% of cases. The modeshapes of the generator speeds for the unstable mode with eigenvalue $0.0061 \pm 2.3740j$ is estimated by BEM-PC, and is shown in Fig.~\ref{fig:mode_shapes_unstable_NENY}(a). Clearly, the most poorly-damped mode in the predisturbance condition has become unstable during cascade propagation and generators $G14-G16$ oscillate against the rest of the generators in this mode. Upon prediction of hyperstability, BEM-PC performs a corrective step by tripping line $41-42$ after $5$ s of the latest event using the predefined SPS action, see Fig.~\ref{fig:mode_shapes_unstable_NENY}(b).
}

\begin{figure}[!h]
\centerline{\includegraphics[scale = 0.205, trim= 2.2cm 0cm 3.85cm 0.5cm, clip=true]{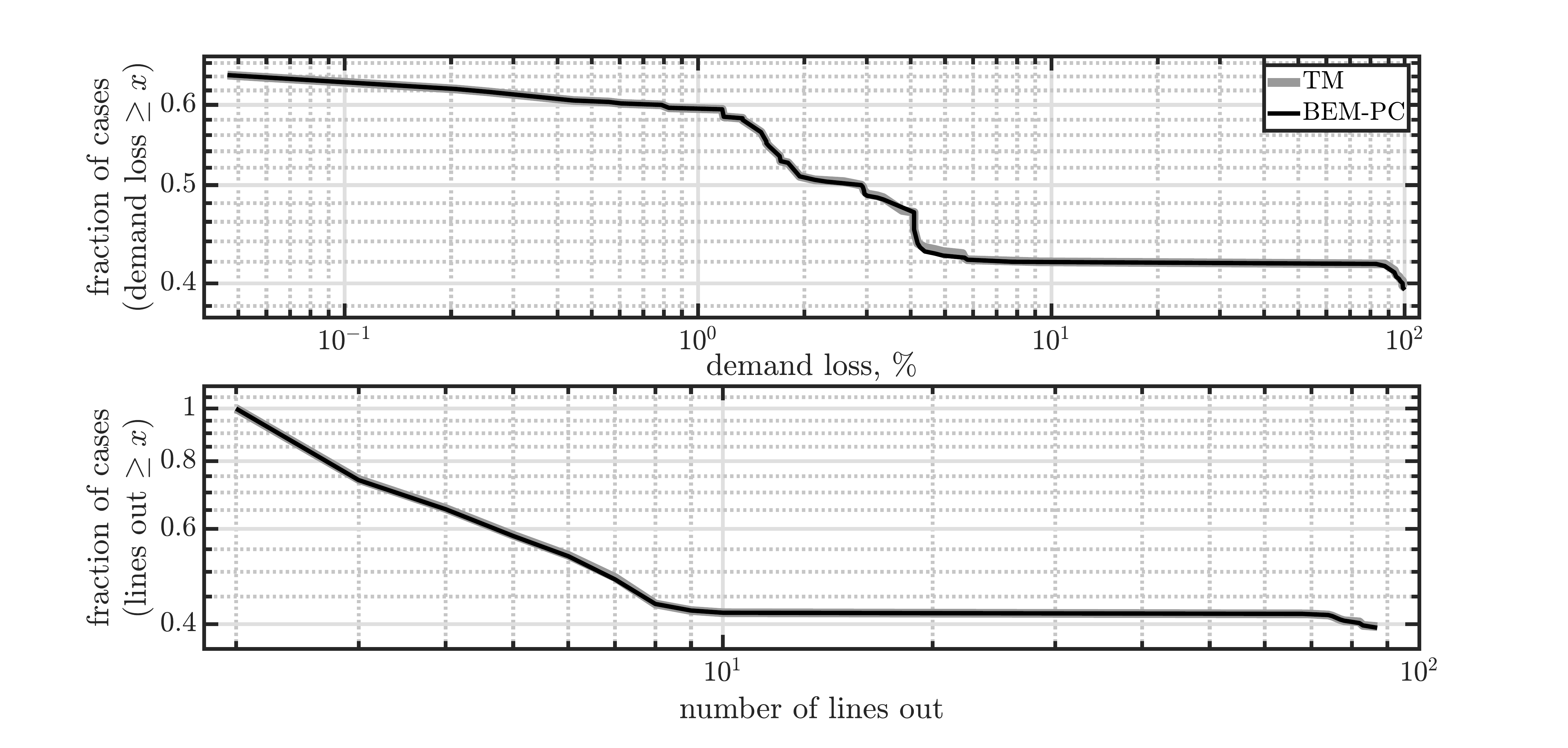}}
\caption{\nrcR{Fraction of cases with $\%$ demand loss $\geq x$ and line outage $\geq x$ at the end of cascade in NE-NY system: comparison between TM and BEM-PC.}}
\centering
\label{fig:BEMPC_TM_NENY_loadline}
\end{figure}

\nrcR{The fraction of cases where the demand loss and line outages at the end of cascade are above a threshold are compared in Fig.~\ref{fig:BEMPC_TM_NENY_loadline}. We see a very close match between TM and BEM-PC. Table VIII shows breakdown of errors in states of buses, machine, and lines at the end of cascade of BEM-PC along with its path agreement measure $R$. It can be seen that the proposed approach is highly accurate in following the actual cascade path, while achieving $\approx 20\times$ speedup on an average, in spite of naturally occurring hyperstability problem.}



\begin{table}[!h]
\centering
\label{tab:number-non-zero-err-cases-NENY} 
\caption { Number of cases with/without hyperstability detection by BEM-PC during cascade: NE-NY system}
\begin{tabular}{ccc}\hline\hline
\multirow{2}{*}{} & \# of cases with & \# of cases without \\
 & hyperstability & hyperstability \\\hline
number & 82 & 418 \\\hline
percentage, \% & 16.4 & 83.6\\\hline
\end{tabular}
\end{table}
\vspace{-0.6cm}

\begin{table}[!h]
\centering
\label{tab:detailed-comp-BEMPC-TM-NENY} 
\caption {\nrcR{(a) End of cascade error, (b) path agreement measure, and (c) run time in TM w.r.t. BEM-PC: NE-NY System}}
\begin{tabular}{c|c|c|c|c|c}\hline\hline
\multicolumn{2}{c|}{} & mean & min & max & median \\\hline
error in & buses & 0.132 & 0 & 18 & 0 \\
state of & machines & 0.032 & 0 & 5 & 0 \\
 & lines & 0.138 & 0 & 19 & 0 \\\hline
\multicolumn{2}{c|}{R} & 0.997 & 0.428 & 1 & 1 \\\hline
\multicolumn{2}{c|}{\textbf{runtime ratio}} & \textbf{19.687} & \textbf{0.235} & \textbf{120.737} & \textbf{15.582}\\\hline
\end{tabular}
\end{table}

\begin{table*}[]
\centering
\label{tab:runtime-details_subprocesses} 
\caption {\nrcR{Runtime in seconds of different sub-processes in BEM-PC shown in Fig.~\ref{fig:fig_flow_chart_predictive_corrective}. Case (I): generic case without hyperstability, and Case (II): with hyperstability in Polish system}}
\resizebox{1.4 \columnwidth}{!}{
\begin{tabular}{c|cccc|cccc|c|c}\hline\hline
\multirow{3}{*}{case} & \multicolumn{9}{c|}{BEM-PC} & TM \\\hline
 & \multicolumn{4}{c|}{round 1} & \multicolumn{4}{c|}{round 2} & \multirow{2}{*}{Total} & \multirow{2}{*}{Total} \\\cline{2-9}
 & a & b1 & b2 & b3 & a & b1 & b2 & b3 &  &  \\\hline
I & 829.3 & 276.5 & 25.2 & 28.5 & 0 & 0 & 0 & 0 & 1159.5 & 37615.2 \\
II & 440.2 & 103.9 & 8.1 & 8.3 & 1608.0 & 528.2 & 32.2 & 33.2 & 2762.1 & 111521.0\\\hline
\end{tabular}}
\\ \footnotesize{\textbf{a}: \scriptsize{subprocess (a)}~~~~\textbf{b1}: \scriptsize{equilibrium calculation in subprocess (b)}}\\
\footnotesize{\textbf{b2}: \scriptsize{A matrix calculation attained from (\ref{eq_A_matrix}) in subprocess (b)}~~~~\textbf{b3}: \scriptsize{eigen decomposition in subprocess (b), see Figure \ref{fig:fig_flow_chart_predictive_corrective}}}
\end{table*}
\begin{figure}[!h]
\centerline{\includegraphics[scale = 0.205, trim= 2.2cm 0cm 3.85cm 0.5cm, clip=true]{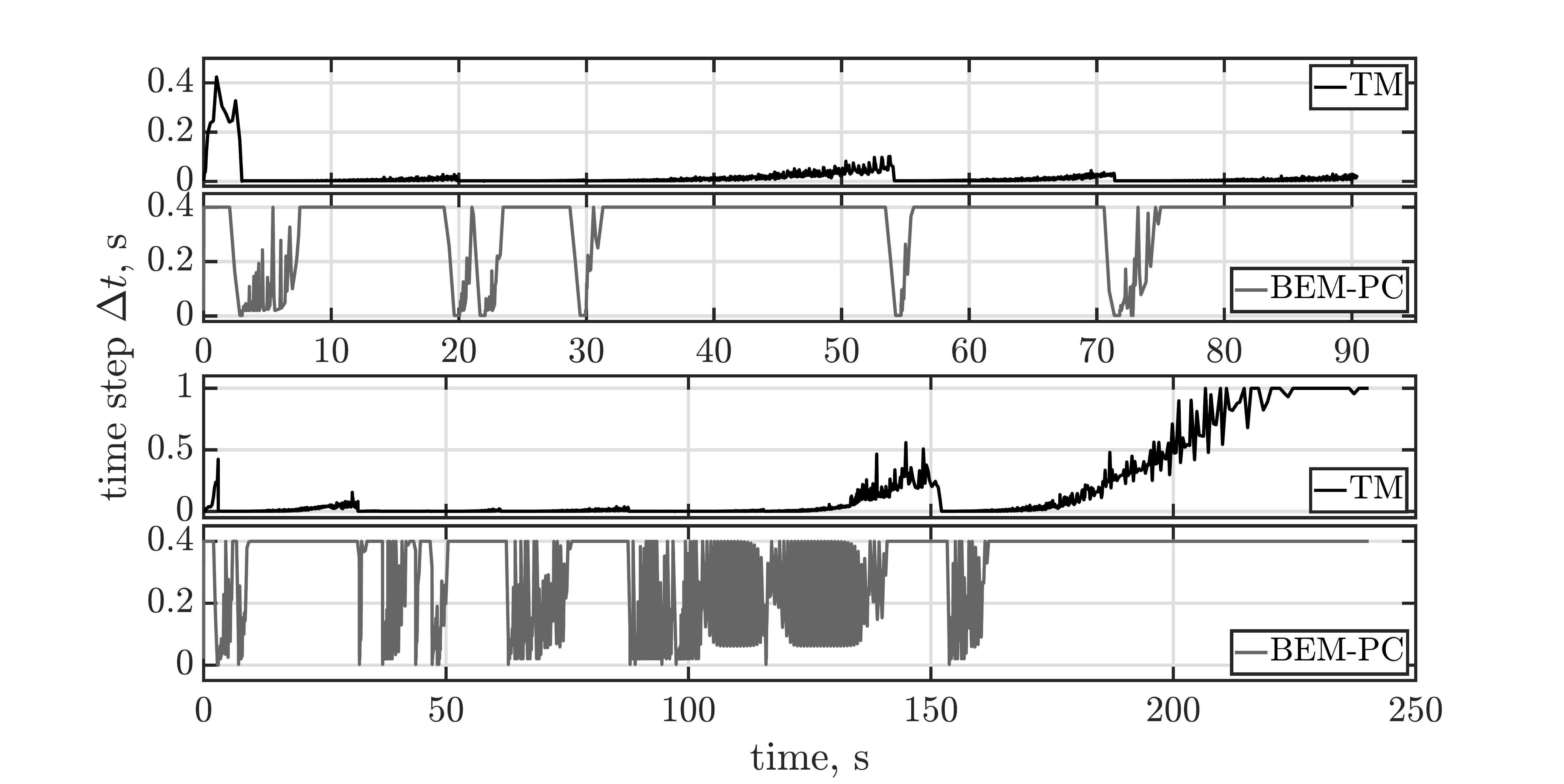}}
\caption{\nrcR{Adaptation of integartion time step $\Delta t$ in TM and BEM-PC. Case (I): Top subplots. Case (II) Bottom subplots. }}
\centering
\label{fig:delta_t_plots_polish_sys}
\end{figure}

\begin{figure}[!h]
\centerline{\includegraphics[scale = 0.205, trim= 2.2cm 0cm 3.85cm 0.5cm, clip=true]{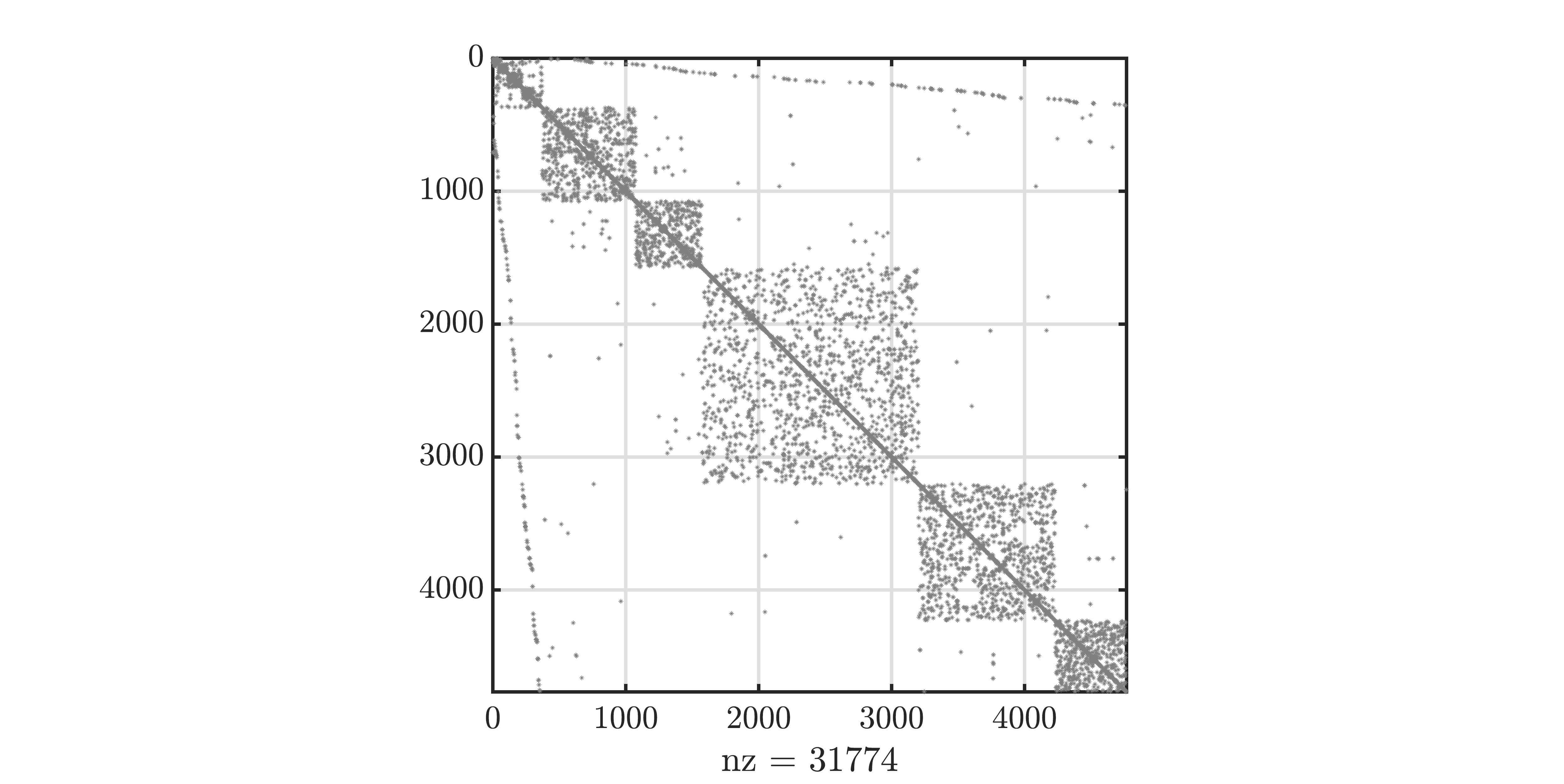}}
\caption{\nrcR{Visualization of sparsity pattern of $J_{22}$ in predisturbance condition with dimension of $4766\times4766$, where $nz$ shows number of nonzero elements. The matrix is $99.86\%$ sparse.}}
\centering
\label{fig:Sparsity_J22}
\end{figure}

\begin{figure}[!h]
\centerline{\includegraphics[scale = 0.205, trim= 2.2cm 2cm 3.85cm 0.5cm, clip=true]{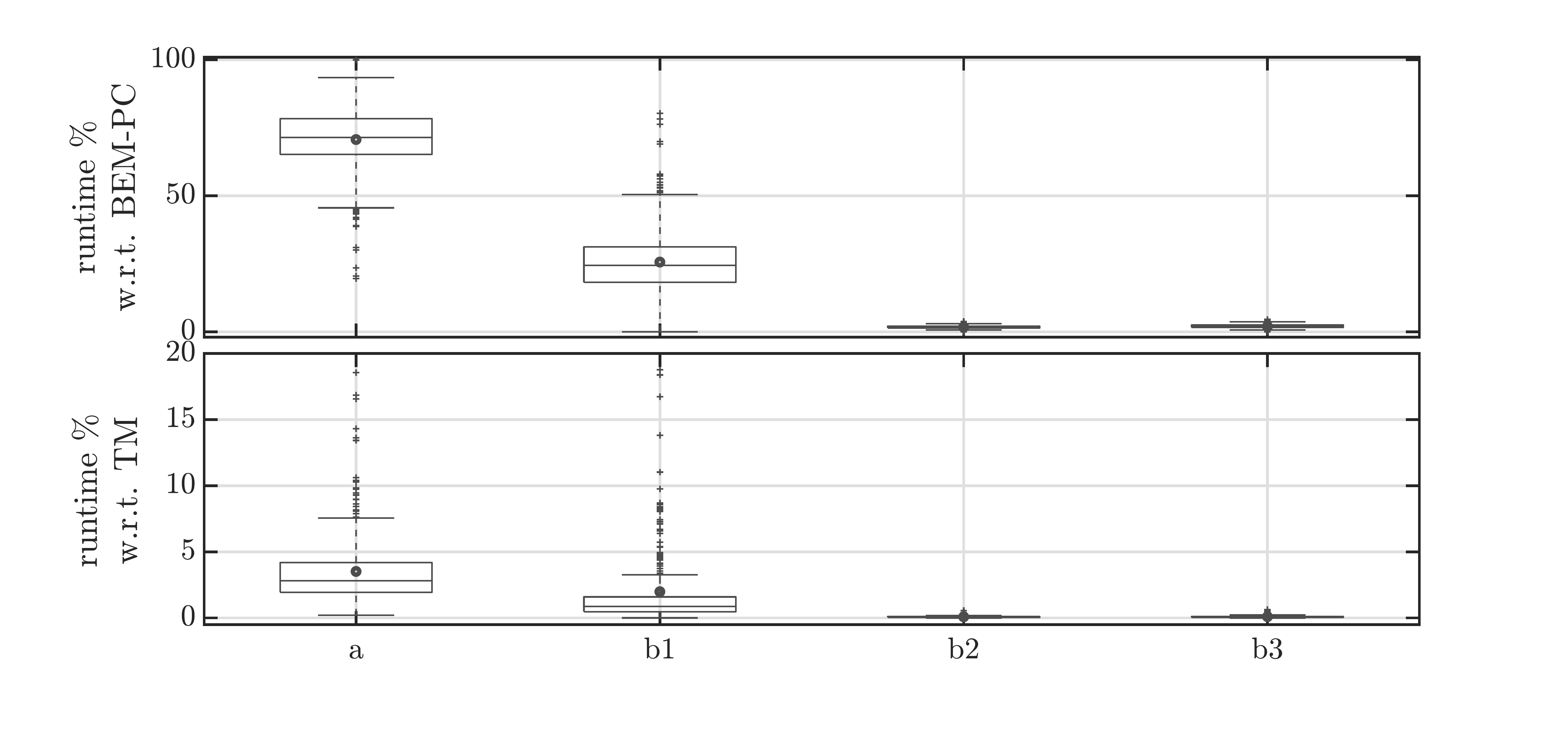}}
\caption{\nrcR{Boxplots of runtimes of different subprocesses of BEM-PC as in Fig. \ref{fig:fig_flow_chart_predictive_corrective} during $500$ MC runs of Polish system. \textbf{a}: subprocess (a). \textbf{b1}: equilibrium calculation in subprocess (b). \textbf{b2}: A matrix calculation from (\ref{eq_A_matrix}) in subprocess (b), and \textbf{b3}: eigendecomposition of A matrix in subprocess (b). Runtimes are expressed as a \% of total runtimes of BEM-PC (top) and TM (bottom).}}
\centering
\label{Fig:sub_process_BEM_runtime}
\end{figure}

\begin{table}[]
\centering
\label{tab:runtime-details_subprocesses_mean_median} 
\caption {\nrcR{Mean and median values of runtimes of different subprocesses of BEM-PC in Fig. \ref{fig:fig_flow_chart_predictive_corrective} as a \% of total runtime of BEM-PC and TM in $500$ MC runs for Polish system}}
\begin{tabular}{c|ccccc}\hline\hline
\multicolumn{2}{c}{\multirow{2}{*}{}} & \multicolumn{4}{c}{BEM subprocess} \\\cline{3-6}
\multicolumn{2}{c}{} & a & b1 & b2 & b3 \\\hline
\multirow{2}{*}{mean} & BEM-PC & 70.69 & 25.62 & 1.71 & 1.98 \\
 & TM & 3.51 & 1.98 & 0.08 & 0.09 \\\hline
\multirow{2}{*}{median} & BEM-PC & 71.42 & 24.41 & 1.7 & 1.94 \\
 & TM & 2.82 & 0.87 & 0.06 & 0.07\\\hline
\end{tabular}
\end{table}

\begin{figure}[!t]
\centerline{\includegraphics[scale = 0.205, trim= 2.2cm 0cm 3.85cm 0.5cm, clip=true]{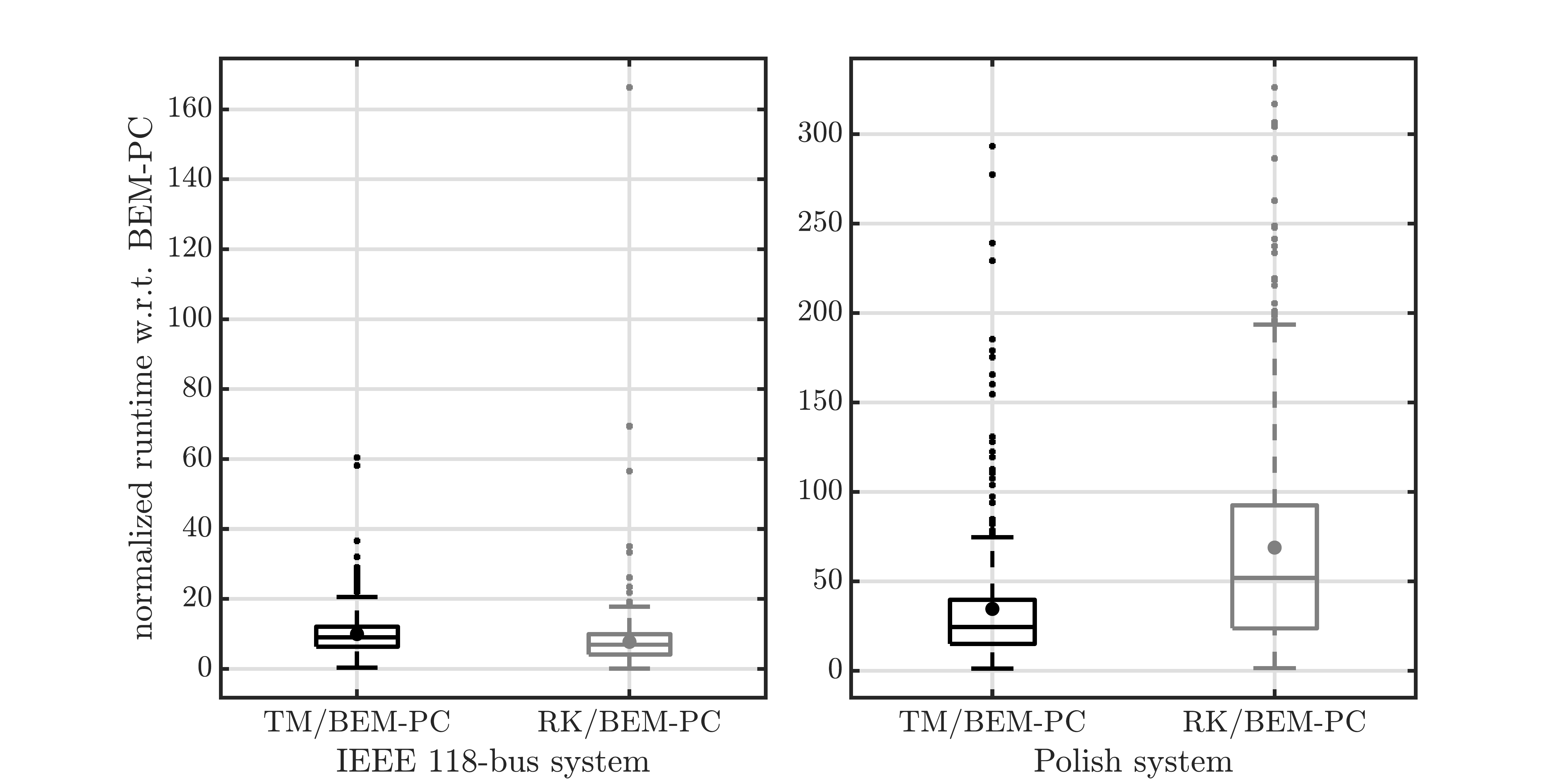}}
\caption{\nrcR{Boxplots of normalized runtime of TM and R-K w.r.t. BEM-PC for cascading failure in -- Left: IEEE 118-bus system, and Right: Polish system.}}
\centering
\label{Fig:boxplot_TM_BEM_RK}
\end{figure}

\vspace{6pt}
\nrcR{\subsection{Analysis of Computational Efficiency}\label{sec:CompuEffNumerical}
In support of the logical arguments presented in Section \ref{sec:CompuEff}, we present our analysis of computational efficiency of BEM-PC based on simulation data. To that end, we performed rigorous data collection relating individual subprocesses in Fig.~\ref{fig:fig_flow_chart_predictive_corrective}. In addition to TM, we also present performance comparison with the partitioned approach widely used in production-grade softwares.
\subsubsection{Performance comparison with TM}\label{sec:CompuEffBEM}
Statistical analysis of the overall CPU time comparison between BEM-PC and TM was presented using the runtime ratio in Tables I-II. The box plots of this metric are also shown in Fig.~\ref{Fig:boxplot_TM_BEM_RK}. Although the  runtime ratio is a good measure of the overall computational efficiency of BEM-PC, it is important to understand how the individual subprocesses in Fig.~\ref{fig:fig_flow_chart_predictive_corrective} contribute towards that. We choose Polish system to demonstrate this due to the scalability challenge it poses. To this end, first we choose two cases -- Case (I): a case without hyperstability, and Case (II): the hyperstability case $\#2$ analyzed in the previous Section. The analysis of BEM-PC subprocesses are perfomed below.\\[3pt]
\textit{1. Subprocess (a):} Figure \ref{fig:delta_t_plots_polish_sys} shows the adaptation of integration time step $\Delta t$ in BEM-PC and TM while solving the cascading process leading to the main surviving island. It can be seen that TM demands much shorter time step whereas BEM-PC enjoys simulation with $\Delta t_{max} = 0.4$ s for most of the simulation period. It can be calculated from Table IX that this subprocess consumes $\approx 70$\% and $\approx 75$\% of BEM-PC's overall CPU time for Cases (I) and (II), respectively. These are however, merely $\approx 2$\% of the TM's runtime in both cases. \\[3pt]
\textit{2. Subprocess (b):} Table IX also shows a breakdown of BEM-PC's runtime within subprocess (b) in Fig.~\ref{fig:fig_flow_chart_predictive_corrective}. As mentioned in Section~\ref{sec:CompuEff}, this subprocess can be segmented further into (b1), (b2), and (b3). Subprocess (b1) is the most expensive among the three, and based on the runtimes shown in Table IX,  it uses $\approx23$\% of BEM-PC's overall CPU time for both cases.
Subprocess (b2) requires inversion of the submatrix $J_{22}$, which is very sparse. To get an idea, Fig.~\ref{fig:Sparsity_J22} shows the sparsity pattern of $J_{22}$ in pre-disturbance condition. The dimension of the matrix is $4766\times4766$, and it is $99.86\%$ sparse. It takes $\approx 2.16$ s to form the corresponding largest $A$ matrix in PSU's ROAR computers \cite{PSU_ROAR} when sparse objects are used in conjunction with Matlab's most comprehensive inversion routine \cite{matlab_mldivide}. The dimension of the $A$ matrix in this case is $1964\times 1964$, and it takes $\approx 2.5$ s for its eigendemposition in subprocess (b3). Based on the results in Table IX, (b2) and (b3) consume $\approx 1.5- 2$\% and $\approx 1.5 - 2.5$\% of total CPU time of BEM-PC, respectively. Note that the subprocess (c) is lookup table-based and consumes negligible CPU time.\\[3pt]
After an in-depth analysis of two specific cases, statistical analysis is performed to assess the computational burden of the subprocesses in $500$ cases of Polish system. Figure~\ref{Fig:sub_process_BEM_runtime} shows the boxplots of these runtimes expressed as percentages of total runtimes of BEM-PC (top) and TM (bottom). The mean and median figures are specified in Table X. The following conclusions can be drawn from these data --
\begin{enumerate}
    \item Subprocess (a) consumes the most significant computational burden followed by (b1). In comparison, both calculation of $A$ matrix and its eigendecomposition requires negligible CPU time.
    \item Aided by the \textit{stiff-decay} property, both of subprocesses (a) and (b1) run significantly faster than TM due to their ability to use larger integration step length $\Delta t$.
\end{enumerate}}

\vspace{-4pt}
\nrcR{
\subsubsection{Performance comparison with partitioned approach}\label{sec:CompuEffRK}
As described in the Introduction section, production grade stability programs use the partitioned approach for dynamic simulation. For a fair comparison, we have performed cascading failure simulations using the partitioned approach with $4^{th}$-order Runge-Kutta \textit{(R-K)} method, which is a widely-used explicit numerical integration technique \cite{kundur_book}. The simulations were run at a fixed time-step of $0.002$ s. Both R-K and TM produces near-identical simulation results. Boxplots of normalized runtime of R-K w.r.t. BEM-PC are shown in Fig. \ref{Fig:boxplot_TM_BEM_RK}. The following are the key observations --
\begin{enumerate}
    \item For the relatively smaller IEEE 118-bus system, R-K is slightly faster than the TM. However, the mean and median ratios of runtime between R-K and BEM-PC from $500$ MC runs are $7.72$ and $6.91$, respectively. For TM these numbers are $9.96$ and $9.05$, respectively.
    \item For the Polish system, R-K is  slower than TM. The mean and median ratios of runtime between R-K and BEM-PC from $393$ MC runs are $68.82$ and $51.84$, respectively. For TM, these numbers corresponding to the same MC runs are $34.54$ and $24.44$, respectively. Note that the remaining $107$ cases of the $500$ MC runs could not be simulated using R-K method as those are running beyond $60$ hours. 
\end{enumerate}
Clearly, BEM-PC retains its advantage over traditional partitioned approach. This is in line with what we mentioned in point $\#5$ of remarks under Section \ref{sec:SimultImplicit}. 
} 
\begin{figure}[!t]
\centerline{\includegraphics[scale = 0.205, trim= 2.2cm 0cm 3.85cm 0.5cm, clip=true]{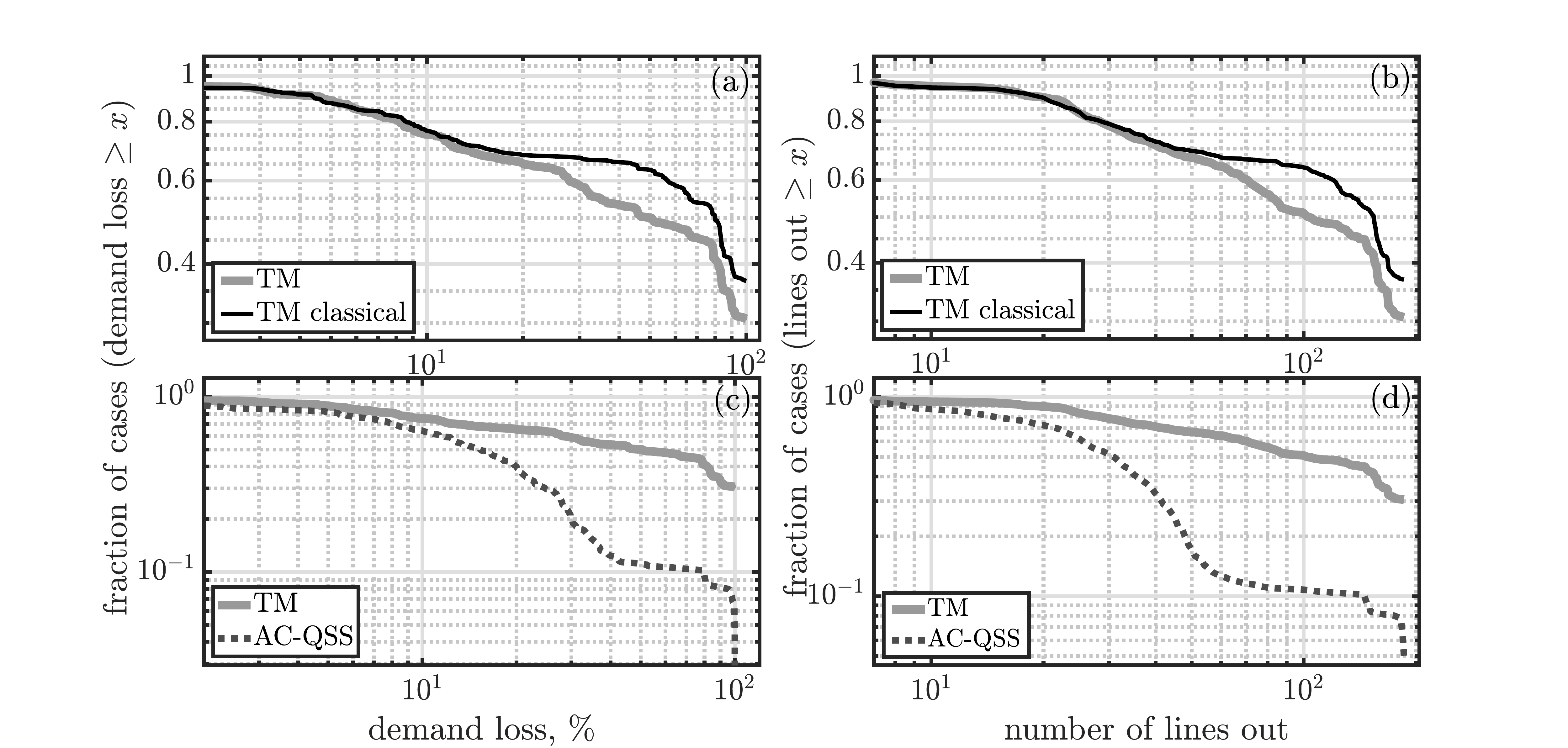}}
\caption{\nrcR{Fraction of cases with $\%$ demand loss $\geq x$ and line outage $\geq x$ at the end of cascade in IEEE 118-bus system: (a),(b)-  comparison between ground truth (TM) and classical models.  (c),(d)- comparison between ground truth (TM) and AC-QSS models.}}
\centering
\label{fig:classic_QSS_118_bus}
\end{figure}
\begin{figure}[!h]
\centerline{\includegraphics[scale = 0.205, trim= 2.2cm 0cm 3.85cm 0.5cm, clip=true]{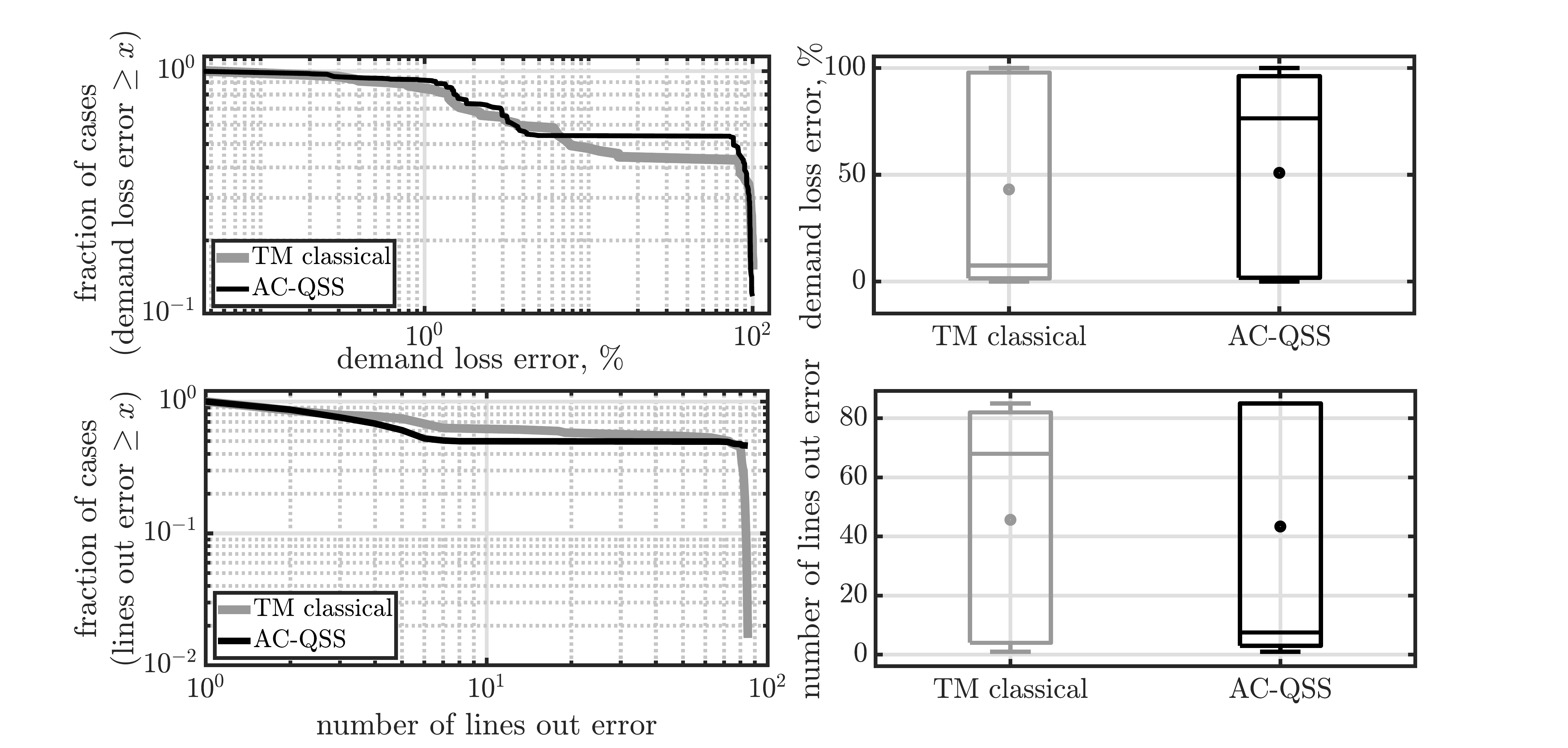}}
\caption{\nrcR{Fraction of cases with $\%$ demand loss error $\geq x$ and line outage error $\geq x$ at the end of cascade in NE-NY system: comparison between ground truth against classical and AC-QSS models for cases with nonzero errors in demand loss and line outage.}}
\centering
\label{fig:TM4th_comparison_TMclassic_QSS_NENY_loadline}
\end{figure}
\begin{table}[!t]
\centering
\label{tab:number-non-zero-err-cases-NENY} 
\caption {\nrcR{Number of cases with nonzero errors in demand loss and line outage at the end of cascade comparing ground truth (TM) against classical and AC-QSS models: NE-NY and Polish systems}}
\begin{tabular}{c|cc|cc}\hline\hline
\multicolumn{1}{c|}{\multirow{2}{*}{}} & \multicolumn{4}{c}{\# of cases with error   in} \\\cline{2-5}
\multicolumn{1}{c|}{} & \multicolumn{2}{c|}{Demand loss} & \multicolumn{2}{c}{Lines out} \\\hline
TM vs & NE-NY & Polish & NE-NY & Polish \\\hline
TM classical & 79 & 76 & 62 & 113 \\
AC-QSS & 297 & 296 & 322 & 342\\\hline
\end{tabular}
\end{table}
\vspace{-0.25cm}
\nrcR{
\subsection{Comparison with AC-QSS and classical models}\label{sec:QCC_Classic_Comp}
In this section, we contrast the cascading failure simulation results in the ground truth (that uses a $4^{th}$-order generator model solved using TM) with the AC-QSS model \cite{Sina-21-TPWRS-UVLS} and dynamic model with classical generator representation. Going forward, \textit{`error'} will imply difference w.r.t. ground truth denoted as `TM.'
\subsubsection{IEEE $118$-Bus System}
Figure~\ref{fig:classic_QSS_118_bus} shows that both models demonstrate similarity with the $4^{th}$-order model, when cascading failure is less severe. However, as the cascade leads to further line outages and load tripping, the AC-QSS and the classical model start departing from the ground truth. For this system, the AC-QSS model shows an optimistic result, while the classical model presents a pessimistic result, when compared with the ground truth.} 

\nrcR{
\subsubsection{IEEE $68$-Bus NE-NY System}
We first look into the modal characteristics of the system before initial outages. The modeshapes of generator speeds corresponding to the most poorly-damped modes are shown in Fig.~\ref{fig:mode_shapes_predis_NENY} for $4^{th}$-order synchronous generator representation vs classical model representation. Next, Table XI shows the number of cases with nonzero error with respect to ground truth in line outage and demand served at the end of cascade when classical generator-based model and AC-QSS model are used in NE-NY system. Finally, figure~\ref{fig:TM4th_comparison_TMclassic_QSS_NENY_loadline} quantifies the error in such cases. The following are the key observations from Figs~\ref{fig:mode_shapes_predis_NENY},~\ref{fig:TM4th_comparison_TMclassic_QSS_NENY_loadline}, and Table XI --
\begin{itemize}
    \item The modal characteristics of the $4^{th}$-order and classical model-based systems are quite different. For the former, the most poorly-damped mode is $-0.0804 \pm 2.4474j$, whereas for the latter, it is $-0.0718 \pm 5.0125j$. In $4^{th}$-order model, generators $G14-16$ oscillate against those in NETS and NYPS for this mode, whereas for the classical model, $G15$ oscillates against $G14$ and $G16$ for the corresponding mode.
    \item Figure~\ref{fig:TM4th_comparison_TMclassic_QSS_NENY_loadline} reveals that in the classical model among the cases in Table XI the mean demand loss error is higher than $40$\% and the mean line outage error is more than $40$. These errors are similar for the AC-QSS model, but for a much larger number of cases, as shown in Table XI.
\end{itemize}
}
\vspace{-.2cm}

\begin{figure}[!t]
\centerline{\includegraphics[scale = 0.205, trim= 2.2cm 0cm 3.85cm 0.5cm, clip=true]{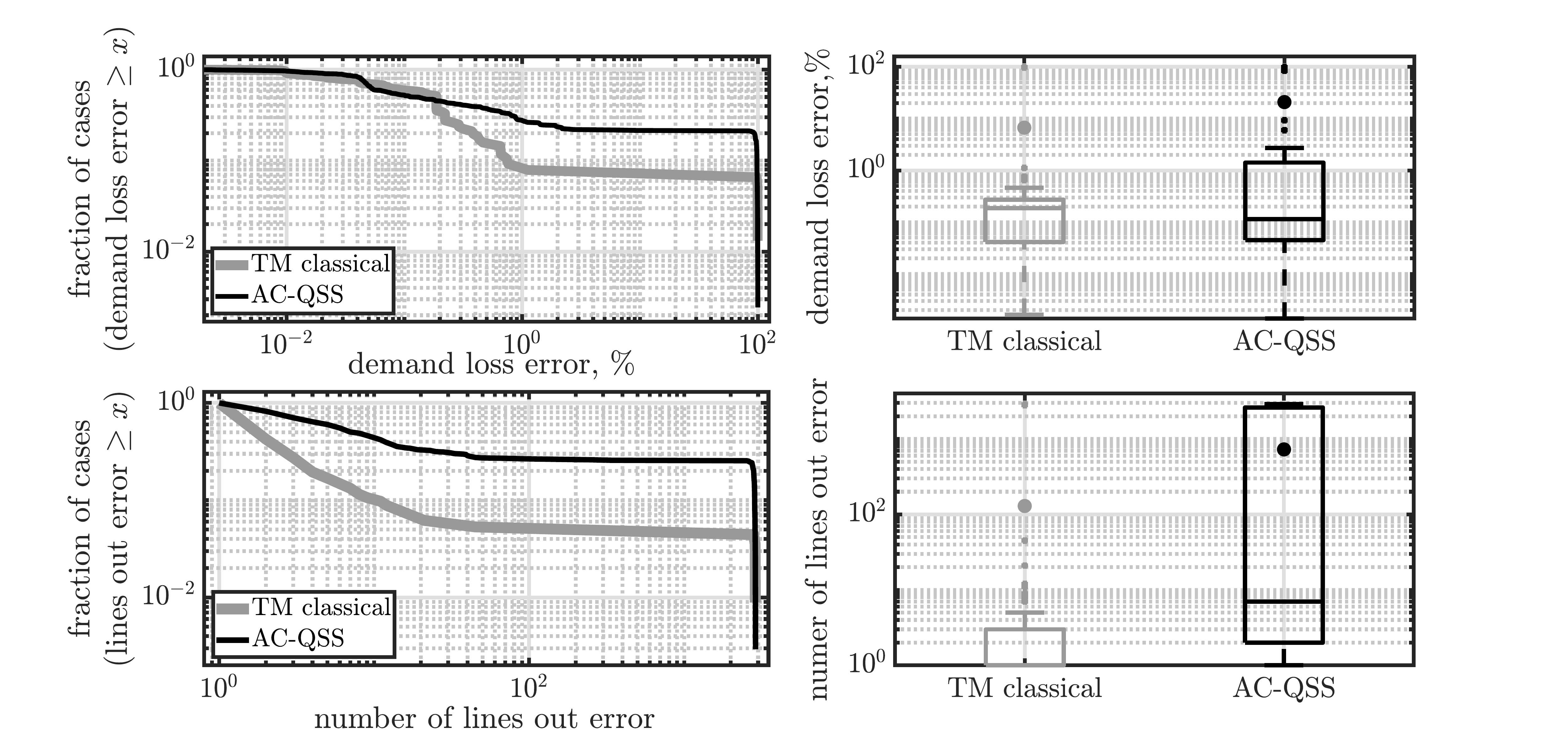}}
\caption{\nrcR{Fraction of cases with $\%$ demand loss error $\geq x$ and line outage error $\geq x$ at the end of cascade in Polish system: comparison between ground truth against classical and AC-QSS models for cases with nonzero errors in demand loss and line outage.}}
\centering
\label{fig:TM4th_comparison_TMclassic_QSS_Polish_loadline}
\end{figure}

\nrcR{
\subsubsection{Polish System}
Table XI shows the number of cases with nonzero error with respect to ground truth in line outage and demand served at the end of cascade when classical generator-based model and AC-QSS model are used in Polish system. Figure~\ref{fig:TM4th_comparison_TMclassic_QSS_Polish_loadline} quantifies the error in such cases. In line with the expectations, the AC-QSS model shows higher error than the classical model.
}

\vspace{-4pt}
\section{Conclusion and Future Work} 
\nrc{A fast time-domain cascading failure simulation approach based on implicit Backward Euler method (BEM) with stiff decay property is proposed in this work. To solve the hyperstability problem of BEM, we proposed a parallelizable predictor-corrector (BEM-PC) approach requiring eigendecomposition of the system matrix corresponding to the linear model obtained around the post-event unstable equilibrium, which BEM converges to. The system matrix is obtained as a by-product of BEM. The proposed BEM-PC approach is benchmarked in a serial implementation against the traditional Trapezoidal method (TM)-based approach. It has shown on an average $\approx 10\times$ speedup in IEEE 118-bus system, \nrcR{$\approx 20\times$ speedup in IEEE 68-bus system,} and $\approx 35\times$ speedup in the Polish grid based on $500$ simulations in each system with random node outages while following exact cascade paths and end results as in TM in most of the cases. \nrcR{It was also shown that BEM-PC retains its computational advantage with respect to partitioned approach using Runge-Kutta-based numerical integration method. Finally, it was shown that AC-Quasi-Steady-State and classical generator model-based representations can lead to different results when compared with a detailed model with $4^{th}$-order generator and exciter dynamics.} Our ongoing and future work focuses on parallelization of BEM-PC, which should lead to further speedup.}

\renewcommand*{\bibfont}{\footnotesize}

\printbibliography

\end{document}